\documentclass[12pt,showpacs,preprintnumbers,superscriptaddress,epsf,amsmath,amsfonts,amssymb,latexsym]{article}

\usepackage{graphics,graphicx}
\usepackage{cite,color,bm}

\input{colordvi.tex}
\renewcommand{\em}{}

\def\ga{\hspace{0.3em}\raisebox{0.4ex}{$>$}\hspace{-0.75em}\raisebox{-.7ex}{$\sim$}\hspace{0.1em}}

\def\gtrsim{\ga}

\newcommand{\eqref}[1]{(\ref{#1})}

\def\l{\ell}
\def\gi{{\rm g}_i}

\def\yi{\gamma_i}

\def\fsky{f_{\rm sky}}
\def\ng{\bar{n}_{\rm g}}
\def\lmin{\ell_{\rm min}}
\def\lmax{\ell_{\rm max}}

\def\nhat{\hat{\bf n}}
\def\khat{\hat{\bf k}}

\def\nn{\nonumber}

\def\LBCMB{~~SNe + CMB}
\def\LBALL{~~SNe + CMB + LSS}

\newcommand{\prd}{Phys. Rev. D}
\newcommand{\apj}{ApJ}
\newcommand{\physrep}{Physics Report}
\newcommand{\mnras}{MNRAS}
\newcommand{\jcap}{JCAP}

\begin{document}

\setcounter{footnote}{0}

\begin{titlepage}

\begin{center}

\vskip .15in

{\Large \bf
Probing Quintessence Potential \\ with Future Cosmological Surveys
}

\vskip .45in

{\large
Yoshitaka Takeuchi$\,^1$, 
Kiyotomo Ichiki$\,^2$,
Tomo~Takahashi$\,^3$,  \\
and 
Masahide Yamaguchi$\,^4$
}

\vskip .45in

{\it
$^1$
Department of Physics and Astrophysics, 
Nagoya University, Nagoya 464-8602, Japan \vspace{5pt} \\
$^2$
Kobayashi-Maskawa Institute for the origin of particles and the
universe, Nagoya University, Nagoya 464-8602, Japan \vspace{5pt} \\
$^3$
Department of Physics, Saga University, Saga 840-8502, Japan \vspace{5pt} \\
$^4$
Department of Physics, Tokyo Institute of Technology,
Tokyo 152-8551, Japan
}

\end{center}

\vskip .4in


\begin{abstract}
Quintessence, a scalar field model, has been proposed to account for
the acceleration of the Universe at present. We discuss how accurately
quintessence models are discriminated by future cosmological surveys,
which include experiments of CMB, galaxy clustering, weak lensing, and
the type~Ia SNe surveys, by making use of the conventional
parameterized dark energy models.  We can see clear differences
between the thawing and the freezing quintessence models at more than
$1\sigma$ ($2\sigma$) confidence level as long as the present equation
of state for quintessence is away from $-1$ as $w_X \gtrsim
-0.95~(-0.90)$. However, it is found to be difficult to probe the
effective mass squared for the potential in thawing models, whose
signs are different between the quadratic and the cosine-type
potentials. This fact may require us to invent a new estimator to
distinguish quintessence models beyond the thawing and the freezing
ones.
\end{abstract}

\end{titlepage}

\renewcommand{\thepage}{\arabic{page}}
\setcounter{page}{1}
\renewcommand{\thefootnote}{\#\arabic{footnote}}


\section{Introduction}
\label{sec:intro}

Current cosmological observations indicate that the Universe is
currently in an accelerating stage, which suggests that the present
Universe is dominated by an enigmatic component called dark energy.
Although current cosmological observations are so precise that we can
have some information on its properties such as its current energy
density, the equation of state and so on, we still do not know what the
dark energy is. Since the identification of dark energy is one of the
most important problems in contemporary science, many researches have
been performed to pursue this issue and a lot of models for dark energy
have been proposed (for early works, see, e.g., \cite{quint,quint2,quint3,Chiba:2000,Kamenshchik:2001}).

Among various dark energy models, quintessence, a scalar field model
with the canonical kinetic term, may be the simplest dynamical
dark energy model \cite{quint,quint2,quint3}. The presence of such a
scalar field might be well motivated by theories of particle physics
such as supergravity and/or superstring, which include scalar fields
with a variety of types of the potential. However, the form of potential
is strongly constrained to realize the current accelerating expansion
and the proposed quintessence potentials can be roughly divided into two
types \cite{Caldwell:2005tm}, depending on how a quintessence
field evolves: freezing (tracker) and thawing types.  For the freezing
(tracker) type, a quintessence field moves fast in the early Universe,
then ``freezes'' at some later time to realize its equation of state
$w_X$ close to $-1$. A typical example of potentials in this category is
the inverse power law potential
\cite{quint2,tracker_quin,tracker_quin_model}. For such a potential, it
is well known that the quintessence field exhibits a tracking behavior
where it traces the equation of state of the dominant background fluid
(e.g., radiation and matter). This is the reason why the freezing type
is sometimes called ``tracker model''. On the other hand, in the thawing
model, a quintessence field almost stays somewhere on the potential in
the early times, during which it behaves like a cosmological
constant. At later times, the quintessence field starts to ``thaw'' or
move, that is, starts to roll down the potential \cite{Caldwell:2005tm}.
The situation is quite similar to that of the inflationary dynamics. As
in the case of inflation, typical potentials of this type are (positive)
power law and hilltop type \cite{pNGB_quin} potentials as in chaotic and
new inflation models, respectively.  The difference of these potentials
is characterized by the sign of the effective mass squared.

In light of this consideration, if we limit ourselves to a quintessence
model as dark energy, the first thing we have to do is to differentiate
these two types, i.e., {\it freezing} or {\it thawing}, by using
cosmological observations. Unfortunately, current observational
constraints are not precise enough to discriminate these types.
However, future experiments may well probe the differences and
distinguish these models, which is one of the issues that we are going
to address in this paper.  For this purpose, we investigate constraints
expected from future surveys of cosmic microwave background (CMB),
galaxy clustering and galaxy lensing shear by using their auto- and
cross-correlations. We also make use of future observations of type Ia
supernovae (SNeIa).

In this paper we adopt typical potentials as the fiducial models of the
two types of quintessence. Specifically, we consider a power-law potential
for the freezing type while the quadratic and the cosine-type potentials
for the thawing type models to generate mock data for future
surveys. Then, the generated mock data are fitted to some
phenomenological dark energy models with its equation of state $w_X$
being parameterized in a simple form, which have been adopted in many
works. Our primary goal is, therefore, to investigate how the confidence
regions look like on the plane of dark energy parameters, and to clarify
whether one can differentiate the two types  from future cosmological surveys.

If one can pin down the type of quintessence, i.e., freezing or thawing type,  
the next task is to extract the information of the
potential in more detail. In particular, for the thawing type, we have
two different, but representative examples of the potential such as
quadratic and cosine-type potentials. One of the differences between the
two is characterized by the sign of the effective mass squared. This
difference is important from both theoretical and observational
reasons. From the model building point of view, it is challenging to
construct a quadratic potential because it must be flat beyond the
Planck scale in order to dominate the energy density at late times.
However, once such a potential is realized, the field can easily take a
suitable value during inflation \cite{Ringeval:2010hf}. On the other
hand, the hilltop (cosine) type potential \cite{pNGB_quin} naturally
appears as an axion potential from particle physics models. For this
type of potential, the field value is not necessarily required to exceed
the Planck scale thanks to the offset of the potential, although it
suffers from a severe initial value problem because there is no natural
reason to keep sitting on the top of the potential. Thus, it is quite
important to distinguish the sign of the effective mass squared of the
potentials from this viewpoint. In fact, the sign of the effective mass
squared appears in the evolution equation of perturbations so that it
may leave significant imprints on the large scale structure (LSS). Thus,
we also investigate how accurately we can determine the effective mass
squared of the potentials from the future observational data. For this
purpose, we introduce a new parameterization for dark energy equation of
state, in which the sign of the effective mass squared is incorporated
as a free parameter.

The organization of this paper is as follows. We briefly summarize the
quintessence potential for fiducial models adopted in our analyses in
Sec.~\ref{sec:quin}, and some phenomenological parameterizations for
the dark energy equation of state are presented in Sec.~\ref{sec:DEparam}.
In Sec.~\ref{sec:perturbation}, we summarize the procedure of our
analyses.  The results of the analyses are given in
Sec.~\ref{sec:forecast}.  The final section is devoted to the conclusion
and summary of this paper.

\section{Quintessence Scalar Field and its models}
\label{sec:quin}

The equation of motion for a scalar field $\phi$ in a flat Universe is
given by
\begin{equation}
    {\phi}'' + 2{\cal H}{\phi}' + a^2 V_{,\phi}(\phi) = 0 ,
\label{eq:EoMphi}
\end{equation}
where primes denote the derivative with respect to the conformal time,
${\cal H}={a}'/a$ is the conformal Hubble parameter, 
$V(\phi)$ is a potential of scalar field and $V_{,\phi} = dV/d\phi$.  
The energy density and pressure of a scalar field are given by 
\begin{eqnarray}
 \rho_\phi &=& \frac{1}{2a^2}\phi'^2 + V(\phi)~, \\
 p_\phi &=& \frac{1}{2a^2}\phi'^2 - V(\phi)~,
\end{eqnarray}
respectively. The ratio of the two, $w_X=p_\phi/\rho_\phi$, is an
equation of state for a scalar field. Since we use cosmological data
including the information of density fluctuations such as CMB, one also
has to consider fluctuations in a quintessence field.  Working in the
synchronous gauge, the evolution equation for fluctuations of the
quintessence field is given by
\begin{equation}
 {\delta\phi}''+2{\cal H}{\delta\phi}'
  +\frac{k^2}{a^2}\delta\phi
  +\frac{1}{2}{h}' \phi'
  +a^2 V_{,\phi\phi}\delta\phi=0~,
\end{equation}
where $h$ is metric perturbation in the synchronous gauge \cite{Ma:1995}
and $V_{,\phi\phi}=\frac{d^2 V}{d\phi^2}$. Fluctuations of density,
pressure and velocity divergence of quintessence are, respectively, given
by
\begin{eqnarray}
 \delta\rho_\phi &=& \frac{1}{a^2}\phi'{\delta\phi}'
  +V_{,\phi}\delta\phi~, \\
 \delta p_\phi &=& \frac{1}{a^2}\phi'{\delta\phi}'
  -V_{,\phi}\delta\phi~, \\
 \theta_\phi &=& k^2 \frac{\delta\phi}{{\phi}'}~.
\end{eqnarray}

As mentioned in the introduction, quintessence models can be divided
into two types, i.e., thawing and freezing (tracker) ones. Although there
are still some variations of the potential form for each type, as
fiducial models in this paper, we adopt the following representative
potentials discussed in the literature.

For the thawing quintessence model, only two types of potentials have
been considered in most cases. One is a simple quadratic potential,
whose explicit form is
\begin{equation}
\label{eq:V_phi2}
V(\phi)  = \frac{1}{2} m_\phi^2 \phi^2,
\end{equation}
where $m_\phi$ is the mass for the quintessence.  Since $\phi$ begins to roll
down the potential at very late epoch to realize the accelerating
expansion today, the mass is roughly of the order of the present Hubble
parameter $H_0$. In addition, in order to dominate the energy density at
late time, the field value $\phi$ needs to be close to the Planck value.

The other type is the cosine one of the form
\cite{pNGB_quin}:
\begin{equation}
\label{eq:V_cosine}
V(\phi)  = M^4 \left(  1 - \cos ( \phi / f_\phi) \right).
\end{equation}
This pseudo-Nambu-Goldstone boson type of potential can well be
motivated from some particle physics models and has been discussed in
many papers (see, e.g., \cite{pNGB_quin}).

Concerning the freezing (tracker) type, although many forms of potential
have been argued, most of them include inverse power law or exponential
parts in the potential, which allow an attractor solution. Here we
consider a typical inverse power law type of the form:
\begin{equation}
\label{eq:V_tracker}
V(\phi)  = \beta M_{\rm pl}^4 \left( \frac{\phi}{M_{\rm pl}} \right)^\alpha.
\end{equation}
As mentioned above, there have been many models (potentials) discussed
so far.  For other freezing (tracker) types, see e.g.,
\cite{tracker_quin}. In this paper, we take these potentials as the fiducial
models for each type.

\section{Dark energy parameterizations}
\label{sec:DEparam}

Since we have not yet understood the nature of the dark energy, it is
common to treat dark energy as an ideal fluid with a negative equation
of state. Here we follow this procedure and investigate how dark energy
parameters, in particular, its equation of state defined
below are constrained from future surveys. Although various
parameterizations for $w_X$ can be found in the literature, 
we assume three parameterizations as described below.

\subsection{Parameterization I \& II}
\label{sec:BackgroundEvo}

One of the aims of this paper is to discriminate the freezing and the
thawing models, whose equation of states decreases and increases with
time, respectively.  This feature can be captured by the following
parameterization \cite{Chevallier:2000qy,Linder:2002et}, whose form is
given by
\begin{equation}
w_X (a) = w_0 + (1-a) w_1 = w_0 + \frac{z}{1+z} w_1 \, \qquad {\rm (Parameterization~I)}. 
\label{eq:eos1}
\end{equation}
It is expected that $w_1 < 0$ for the freezing model while $w_1 > 0$ for
the thawing model. We call this parameterization ``Parameterization I''
in the following. With this equation of state $w_X$, one can write its energy
density as
\begin{equation}
\rho_X(z) = \rho_{X0}  (1+z)^{3(1+w_0+w_1)}  \exp \left( \frac{-3 w_1 z}{1+z} \right),
\end{equation}
where $\rho_{X0}$ is the energy density of dark energy at present
time. In the above parameterization, $w_X$ has a simple linear
dependence on the scale factor $a$, but another parameterization of this
type is also possible such as $w_X = w_0 + w_1 z$, where a linear
dependence on a redshift $z$ is assumed
\cite{Huterer:2000mj,Weller:2001gf,Frampton:2002vv} with a cutoff at
some redshift to avoid large $w_X$ at earlier time.  But in the
following, we consider the parameterization of Eq.~\eqref{eq:eos1} as a
representative one of this category.

The equations of states for the freezing and the thawing models are also
characterized by the fact that $w_X$ is larger (smaller) in the past for
the freezing (thawing) model. This feature can be directly probed by the
following parameterization \cite{Hannestad:2004cb},
\begin{equation}
w_X (a) 
= 
w_a w_b \frac{a^q + a_s^q}{w_b a^q + w_a a_s^q}  
=
w_a w_b \frac{1+ \left( \frac{1+z}{1+z_s}\right)^q}{w_b + w_a \left( \frac{1+z}{1+z_s}\right)^q}  
\qquad {\rm (Parameterization~II)},
\label{eq:eos2}
\end{equation}
where the equation of state changes from $w_b$ to $w_a$ at the
transition redshift $z_s$ with its width characterized by $q$. It is
expected that $w_b > w_a$ for the freezing model while $w_b < w_a$ for
the thawing model. We call this form of $w_X$ ``Parameterization II'' in
this paper. The energy density of dark energy with this parameterization
can be analytically given by
\begin{equation}
\rho_X(z) 
= 
\rho_{X0} (1+z)^{3 ( 1+{w_a})} 
\left(  
\frac{w_a + w_b (1+z_s)^q}{w_a (1+z)^q + w_b (1+z_s)^q} 
\right)^{\frac{3({w_a-w_b})}{q}}.
\end{equation}

\subsection{Perturbation evolution and Parameterization III}
\label{sec:perturbEvo}

The above two parameterizations are suitable for discriminating the
thawing and the freezing models. Once the thawing models would be
favored from future observations, the next task is to differentiate 
two typical thawing models, in which the signs of the effective mass
squared are opposite, that is, positive and negative. Such a difference
in the sign of the effective mass squared is expected to affect the
evolution of the perturbations. Thus, in this subsection, we discuss what
kind of parameterization is suitable for probing the sign of the
effective mass squared of a quintessence field through its
perturbations.

For this purpose, let us first remind ourselves how fluctuations in
dark energy fluid evolve with a given equation of state $w_X$. Working
in the synchronous gauge \cite{Ma:1995}, the energy density and
velocity perturbation evolutions for a general fluid with its equation
of state $w_X$ in the CDM rest frame are given by
\begin{eqnarray}
  {\delta}' &=& -(1+w_X)\left( \theta + \frac{{h}'}{2} \right)
  -9{\cal H}^2(1+w_X)(c_{\rm s}^2 - c_{\rm a}^2)\frac{\theta}{k^2}
  - 3 {\cal H} (c_{\rm s}^2 -w_X) \delta ~,
  \label{eq:deltadot} \\
  {\theta}' &=& - {\cal H} (1-3c_{\rm s}^2) \theta 
  + \frac{c_{\rm s}^2}{(1+w_X)} k^2 \delta - k^2 \sigma ~,
  \label{eq:thetadot}
\end{eqnarray}
where $\delta$ and $\theta$ represent density and velocity
perturbations, respectively, and $\sigma$ is anisotropic stress.  One
can consider non-vanishing $\sigma$ for a general dark energy fluid
\cite{Ichiki:2007}, but we can safely set $\sigma=0$ for a quintessence
field. Furthermore, the effective sound speed $c_s$, which is the sound
speed defined in the dark energy rest frame, is defined through
\cite{Hu:1998}
\begin{equation}
  \frac{\delta p}{\rho} = c_s^2 \delta 
  + 3{\cal H}(1+w_X)(c_s^2-c_a^2)\frac{\theta}{k^2} ~ .
\end{equation}
For a scalar field with the canonical kinetic term, the effective
sound speed is $c_s^2=1$.  The adiabatic speed of sound $c_a$ for a
fluid is given by
\begin{equation}
  c_{\rm a}^2 \equiv 
  \frac{{p}'}{{\rho}'} 
  = w_X - \frac{{w_X}'}{3{\cal H}(1+w_X)} ~.
\label{eq:cs2adi}
\end{equation}

By analogy with the case of inflation, let us try to parameterize the
equation of state by using its potential, which is characterized by 
the slow-roll parameters \cite{Ilic:2010,Chiba:2009}. The adiabatic
sound speed Eq.~(\ref{eq:cs2adi}) can be written with the scalar field
$\phi$ and the potential $V$ as
\begin{equation}
    c_{\rm a}^2 = \frac{{p}_\phi'}{{\rho}_\phi'} = \frac{\dot{p}_\phi}{\dot{\rho}_\phi}
      = \frac{\ddot{\phi} - V_{,\phi}}{\ddot{\phi} + V_{,\phi}} 
      = -1 +2\frac{\ddot{\phi}}{\ddot{\phi} + V_{,\phi}} ~,
\label{eq:cs2_phi}
\end{equation}
where dots denote the derivative with respect to the cosmic time.
We first introduce the following function \cite{Crittenden:2007,Linder:2006} 
\begin{equation}
    \beta = \frac{\ddot{\phi}}{3{H}\dot{\phi}} ~ .
\end{equation}
Then, using Eq.~(\ref{eq:EoMphi}), $\dot{\phi}$ is written in term of
$\beta$ as
\begin{equation}
    \dot{\phi} = -\frac{V_{,\phi}}{3(1+\beta){H}} ~ .
\label{eq:dotphi}
\end{equation}
When $|\beta| \ll 1$\footnote{ The assumption $|\beta| \ll 1$ is
reasonably satisfied around present time for thawing quintessence models
with quadratic potential, $V=\frac{1}{2}m^2_\phi \phi^2$, or
cosine-type, $V=M(1-\cos(\phi/f))$, (see e.g.,
\cite{Chiba:2009,Chiba:2010}). However, $\beta$ becomes
$\frac{2}{3}$ and $\frac{1}{2}$ in the radiation and matter dominated eras, respectively, for the thawing model. 
}, $\ddot{\phi}$ is given by differentiating the both sides in
Eq.~\eqref{eq:dotphi} with respect to the time as
\begin{equation}
    \ddot{\phi} \simeq
    \frac{V_{,\phi}V_{,\phi\phi}}{9{H}^2} - \frac{(1+w)V_{,\phi}}{2} 
    = \frac{\eta V_{,\phi}}{3} - \frac{(1+w)V_{,\phi}}{2}~,
\label{eq:ddotphi}
\end{equation}
where we have used $\dot{H}/{{H}^2} \simeq - 3(1+w_X)/2$ and
$\eta$ is defined as
\begin{equation}
    \eta \equiv \frac{V_{,\phi\phi}}{3{H}^2} ~,
\label{eq:eta}
\end{equation}
which reduces to the inflationary slow-roll parameter, $\eta =
V_{,\phi\phi}/V$, when ${H}^2 \simeq V/3$.  Using Eqs.~(\ref{eq:eta})
and (\ref{eq:ddotphi}), we can rewrite Eq.~(\ref{eq:cs2_phi}) as\footnote{
We here assume $|1+w_X| \ll 1$ and $|\eta| \ll 1$. The condition $|\eta|
\ll 1$ is valid for slow-roll thawing models. Although this assumption
is not necessarily satisfied similarly as $|\beta| \ll 1$, we here adopt
these assumptions to find a simple parameterization, which accommodate a
possibility to reflect the sign of the effective mass squared.
},
\begin{eqnarray}
    c_{\rm a}^2 
    &\simeq& -1 +2 \left( \frac{\eta}{3} - \frac{1+w_X}{2} \right)
    \left( 1 +\frac{\eta}{3} - \frac{1+w_X}{2} \right)^{-1} , \nn \\
    &\simeq& -1 +2 \left( \frac{\eta}{3} - \frac{1+w_X}{2} \right)
    \left( 1 -\frac{\eta}{3} + \frac{1+w_X}{2} + \cdots \right) , \nn \\
    &\simeq& -2 -w_X +\frac{2}{3}\eta \, .
\label{eq:cs2_eta}
\end{eqnarray} 
From Eqs.~(\ref{eq:cs2adi}) and (\ref{eq:cs2_eta}), we obtain the following
differential equation for $w_X$: 
\begin{equation}
    \frac{dw_X}{d\ln a} = 6(1+w_X)(1+w_X-\frac{1}{3}\eta ) .
\label{eq:dwdlna}
\end{equation}
Assuming $\eta$ is a constant parameter and the present value of the equation of
state is $w_X(a_0) = w_0$, then we can solve Eq.~(\ref{eq:dwdlna}) as
\begin{equation}
    w_X(a) = -1 +\frac{1}{3}\frac{(1+w_0)\eta}
        {(1+w_0) - (1+w_0 - \frac{\eta}{3})a^{2\eta}} 
        \hspace{5mm}{\rm (Parameterization~III)} \, .
\label{eq:eos3}
\end{equation}
We introduce this parameterization, especially, to differentiate the sign
of the effective mass squared and to distinguish the two typical thawing
models.

In the following analysis, we take these three parameterizations
Eqs.~\eqref{eq:eos1}, \eqref{eq:eos2}, and \eqref{eq:eos3} for $w_X$ to
investigate to what extent one can differentiate the potentials of
quintessence.

\section{Analysis}
\label{sec:perturbation}

We here summarize the method of our analysis.  We use angular power spectra from
CMB and large scale structure, taking auto- and cross-correlations among
them to utilize the information thoroughly.

\subsection{Cosmological observables}
\label{sec:observables}

\subsubsection{CMB}

\subsubsection*{Integrated Sachs-Wolfe effect}

The presence of dark energy or the cosmological constant causes the
decay of the gravitational potential, then affects the propagation of
CMB photons along the line of sight. We can observe this effect as a
cumulative one, hence name the (late-time) integrated Sachs-Wolfe (ISW)
effect \cite{Sachs:1967}. Since the cross-correlation between CMB and
LSS such as galaxy clustering and week lensing
field arises due to the ISW effect, the presence of non-vanishing
cross-correlation between CMB and LSS on large
scales can be a signature of dark energy.

CMB temperature fluctuations on large scales mainly come from the
Sachs-Wolfe effect as a primary source and the ISW effect as a secondary
one at late time. Thus we can schematically write the temperature
fluctuation as $\Delta T_{\rm CMB} = \Delta T_{\rm SW} + \Delta T_{\rm
ISW}$. The contribution from the ISW effect $\Delta T_{\rm ISW}$ is
given by
\begin{equation}
    \frac{\Delta T_{\rm ISW}(\nhat)}{T_{\rm CMB}} =  
    \int_0^{\chi_{*}} d\chi \left[ {\Phi}'(\nhat, \chi(z))
                      - {\Psi}'(\nhat, \chi(z)) \right] ,
\end{equation}
where $T_{\rm CMB}$ is the mean temperature of CMB, $\nhat$ is the
direction along the line of sight, $\chi$ is the comoving distance and
$\chi_{*}$ denotes the comoving distance to the last scattering
surface. Here $\Phi$ and $\Psi$ are the curvature perturbation and the
gravitational potential, respectively, which can be related as $\Phi =
-\Psi$ on subhorizon scales, and a prime denotes a derivative with
respect to the conformal time.

Deep in the horizon scale, the gravitational potential can be related to
matter density fluctuations in Fourier space through the Poisson
equation as
\begin{equation}
    \Phi({\bm k} ,z) = 
        - \frac{3}{2} \frac{\Omega_{\rm m,0}H_0^2}{a k^2} \delta({\bm k}, z) .
\end{equation}
On large scales where the the fluctuations are small $\delta({\bm k}, z)
\ll 1$ and the linear theory is valid, they can be written as
$\delta({\bm k}, z) = \delta({\bm k}, 0) D(z)/D_0$, where $D(z)$ is the
linear growth factor normalized as $D(z) \propto a(z)$ in the matter
nominated era and $D_0$ is the value at the present time.

\subsubsection*{CMB lensing potential}

The gravitational potential produced by large-scale structures deflects
CMB photons on the way propagating to us, and produces other
secondary effects on CMB temperature and polarization fields (see,
e.g., \cite{Lewis:2006}). We can understand such effect as a displacement
of patches with deflection angle ${\bm d}(\nhat)$ on the sphere, and the
relationship between the lensed temperature anisotropy
$\tilde{T}(\nhat)$ and the unlensed one $T(\nhat)$ is given by
$\tilde{T}(\nhat) = T(\nhat + {\bm d}(\nhat))$. The deflection angle can
be written with the lensing potential $\psi$ as ${\bm d}(\nhat) = \nabla
\psi(\nhat)$, where $\psi$ is defined as
\begin{equation}
    \psi(\nhat) = -2 \int_0^{\chi_{*}} d\chi
     \frac{\chi_{*} - \chi}{\chi_{*} \chi}
        \Psi(\nhat, \chi(z)) .
\end{equation}
Here we take $\psi(\nhat)$ to be an observable which characterizes weak
lensing effects of CMB.

The lensing potential $\psi$ can be reconstructed from both observed
temperature and polarization fields with a quadratic estimator and the
noise of the lensing potential can be estimated as the reconstruction
error (see, e.g., \cite{Hu:2002,Okamoto:2003,Hirata:2003}).  To compute
the lensing potential ($\psi$), we use a publicly available Boltzmann
code CAMB \cite{Lewis:2000}, and modify it to include a quintessence
scalar field.

\subsubsection{Galaxy Clustering}

We observe the distribution of galaxies as the projected galaxy over-density
in photometric redshift surveys. We consider a tomographic one, in which
we can separate galaxies into some redshift bins. Fluctuations of 
the galaxy distribution in the $i$-th redshift bin are given by
\begin{equation}
    \gi(\nhat) = \int_{z_i}^{z_{i+1}} dz \, b_{\rm g}(z) \, {\cal N}_i(z) \, \delta(\nhat, z) , 
\end{equation}
where $b_{\rm g}(z)$ is the galaxy bias, ${\cal N}_i(z)$ is the
selection function which represents the redshift distribution of sample
galaxies, and the subscript $i$ denotes a redshift bin.

For the redshift distribution of sample galaxies, we adopt the analytic
formula of \cite{Ma:2006}, which includes the effect of photometric
redshift errors as
\begin{equation}
    p_i(z_{\rm ph} | z) = \frac{1}{\sqrt{2\pi}\sigma_z(z)}
        \exp \left[ -\frac{ \left(z-z_{\rm ph}\right)^2}
        {2\sigma_z^2(z)} \right] ,
\end{equation}
where $\sigma_z(z)$ denotes a redshift scatter systematic error and is
given by
\begin{eqnarray}
    \sigma_z(z) = \sigma_z^{(i)}(1+z) , 
\end{eqnarray}
with $\sigma_z^{(i)}=0.03$ for each redshift bin. Then, the selection
function ${\cal N}_i(z)$ which includes the effect of photometric redshift
errors is written as
\begin{equation}
    {\cal N}_i(z) = \int_{z_{\rm ph}^{(i)}}^{z_{\rm ph}^{(i+1)}} dz_{\rm ph}
        N(z) p_i(z_{\rm ph} | z) 
    = \frac{1}{2} \left[ {\rm erf}(x_{i+1}) - {\rm erf}(x_i) \right] N(z) ,
\end{equation}
where $x_i \equiv ( z_{\rm ph}^{(i)} -z )/\sqrt{2}\sigma_z(z)$ and the
redshift distribution of galaxy samples $N(z)$ is assumed to be
\cite{Smail:1994,Amara:2007}
\begin{equation} 
    N(z) \propto z^{\alpha}
        \exp \left[ -\left(\frac{z}{z_0} \right)^{\beta} \right]. 
\end{equation} 
Here $N(z)$ should be normalized as $\int N(z) dz = 1$ and we adopt
$\alpha=2.0$, $\beta=1.5$.  $z_0$ is determined from the relation with
mean redshift $z_{\rm m}$ defined as
\begin{equation}
    z_{\rm m} = \int z N(z) dz .
\end{equation}

We adopt the mass weighted average bias given by
\begin{equation}
    b_{\rm g}(z) = 
        \left. \int_{M_{\rm min}}^{\infty} b_{\rm h}(M,z) \frac{dn_{\rm h}(M,z)}{dM} dM 
        \right/ \left[ \int_{M_{\rm min}}^{\infty} \frac{dn_{\rm h}(M,z)}{dM} dM \right] \, ,
\label{eq:bias}
\end{equation}
where $M_{\rm min}$ is the minimum mass of the halos which host the
galaxies we can observe, $b_{\rm h}(M,z)$ and $dn_{\rm h}(M,z)/dM$
denote the halo bias and the halo mass function, respectively.  We
utilize the models of \cite{Sheth:2001} and \cite{Warren:2006} for the
halo bias and the halo mass function, respectively.  In the following,
we treat $M_{\rm min}$ as a model parameter to determine the galaxy
bias, and also derive its constraint from mock data.

\subsubsection{Galaxy weak lensing}

The light path from distant galaxies is deflected by foreground
large-scale structures.  The effects can be observed as the
magnification or the distortion of images of background galaxies (see,
e.g., \cite{Bartelmann:2001}) and the effects can be evaluated from the
measurements of shear of each galaxy.
Here we consider a galaxy weak lensing survey with photometric
redshift. The averaged galaxy weak lensing shear is given by
\begin{equation}
    \yi(\nhat) = \frac{3\Omega_{\rm m0}H_0^2}{2}
        \int_0^{\infty} d\chi \frac{\chi(z)}{a(z)} 
        \left[ \int_{{\rm max}(z,z_i)}^{z_{i+1}} dz' \, 
        {\cal N}_i(z') \frac{\chi(z') - \chi(z)}{\chi(z')} \right]
         \delta(\nhat, \chi(z))~,
\end{equation}
where we assume a tomographic survey and the subscript $i$ denotes a
redshift bin.

\subsection{Angular power spectra}

An observable given by the projection of the line of sight along the comoving
radial coordinate $\chi$ in direction $\nhat$ is given by
\begin{equation}
    X({\nhat}) = \int d\chi S^X(\nhat\chi, \tau_0-\chi) ,
\end{equation}
where $S^X({\bm k}, \tau)$ represents the source term of $X$.  Assuming
statistical isotropy, angular correlation between two observables $X$
and $Y$ is given by angular power spectrum as
\begin{equation}
    \langle a_{\l m}^X a_{\l'm'}^{Y*} \rangle = C_\l^{XY} 
    \delta_{\l \l'}\delta_{mm'} ,
\end{equation}
where $a_{\l m}^{X}$ are the expansion coefficients of the observables
$X(\nhat)$ in spherical harmonics, $X(\nhat) \equiv \sum_{\l m}a_{\l
m}^{X}Y_{\l m}(\nhat)$, and given by
\begin{equation}
    a_{\l m}^X  = 4\pi\ (-i)^\l \int \frac{d^3k}{(2\pi)^3} Y_{\l m}^{*}(\khat)
      \int d\chi S^X({\bm k},\tau_0-\chi) j_\l(k\chi) .
\end{equation}
If we write the source term as $S^X({\bm k},\tau_0-\chi) =
S^X(k,\tau_0-\chi) \tilde{\Phi}({\bm k})$, then the angular power
spectrum is given by
\begin{equation}
    C_\l^{XY} = \frac{2}{\pi} \int d^3k P_{\Phi}(k) 
      \int d\chi S^X(k,\tau_0-\chi) j_\l(k\chi) 
      \int d\chi' S^Y(k,\tau_0-\chi') j_\l(k\chi') ,
\end{equation}
where $\tilde{\Phi}({\bm k})$ is the Fourier transform of the
primordial curvature perturbation $\Phi(\nhat)$, $P_{\Phi}(k)$ is the
power spectrum of $\Phi$, and $S^X(k,\tau_0-\chi)$ is the transfer
function of $X$.

In the following analysis, we take into account all auto- and
cross-correlations between different observables, except for the
cross-correlations of CMB $E$-mode polarization with CMB lensing, galaxy
clustering and galaxy weak lensing shear. This is because CMB $E$-mode
polarizations are produced at the last scattering surface or at the
epoch of reionization, whose redshifts are much larger than those probed
by CMB lensing, galaxy clustering and galaxy weak lensing.

\subsection{Parameter estimation with angular power spectra}
\label{sec:likelihood}

In order to discuss whether we can probe the differences among
quintessence models, we fit the phenomenological dark energy models
based on the parameterized equation of state given in
Section~\ref{sec:DEparam}, by generating mock data of angular power
spectra ${\bm C}_{\l, {\rm fid}}^{\rm XY}$ for the fiducial quintessence
models, with $X,Y$ representing the observables discussed above.  Following
\cite{dePutter:2010}, we evaluate $\chi^2$ as
\begin{equation}
    \chi^2 = \sum_{{\rm XY}, {\rm ZW}} \sum_{\l} 
        \left( {\bm C}_{\l, {\rm fid}}^{\rm XY} - {\bm C}_{\l, {\rm mod}}^{\rm
        XY}\right)^{\rm T}
        ({{\bf Cov}_{\l}})_{{\rm XY}, {\rm WZ}}^{-1} 
        \left({\bm C}_{\l, {\rm fid}}^{\rm WZ} - {\bm C}_{\l, {\rm mod}}^{\rm WZ} \right) ,
\label{eq:chisq_cls}
\end{equation}
where ${\bm C}_{\l, {\rm mod}}^{\rm XY}$ is the power spectra for the
phenomenological dark energy model and $({{\bf Cov}_{\l}})$ denotes
covariance matrix. Each component of the covariance matrix is given by
\begin{equation}
    {\rm Cov}_\l 
    \left[ C_{\l, {\rm fid}}^{\rm XY}, \, C_{\l, {\rm fid}}^{\rm WZ}\right] 
    = \frac{1}{(2\l + 1)\fsky} 
    \left( \tilde{C}_\l^{\rm XW} \tilde{C}_\l^{\rm YZ}  
        + \tilde{C}_\l^{\rm XZ} \tilde{C}_\l^{\rm YW} \right) ,        
\label{eq:covmat}
\end{equation}
with
\begin{equation}
    \tilde{C}_\l^{\rm XY} \equiv 
        C_{\l, {\rm fid}}^{\rm XY} + N_\l^{\rm XY} .
\end{equation}
Here $\fsky$ is the sky coverage, $N_\l^{\rm X}$ is the noise spectrum
for the auto- or cross-correlations between the observables $X$ and $Y$.
For this noise spectrum, we assume that the correlations of noises
between different observables are negligible; $N_\ell^{\rm
XY}\equiv\delta_{\rm XY}N_\ell^{\rm X}$. \\


\begin{table}[t]
\begin{center}
\begin{tabular}{lcccccc}
\hline \hline
 Experiment & 
 \makebox[1.5cm]{$f_{\rm sky}$} & 
 \makebox[2cm]{$\nu$} & 
 \makebox[2cm]{$\theta_{\rm FWHM}$} & 
 \makebox[2cm]{$\Delta_\nu^T$} & 
 \makebox[2cm]{$\Delta_\nu^P$}  \\
 &  & [GHz] & [arcmin] & [$\mu$K/pixel] & [$\mu$K/pixel]  \\
\hline
COrE  & 0.65 
    & 105 & 10.0' & 0.536  & 0.926  \\
 &  & 135 &  7.8' & 0.674  & 1.167  \\
 &  & 165 &  6.4' & 0.834  & 1.441  \\
 &  & 195 &  5.4' & 0.974  & 1.681  \\
 &  & 225 &  4.7' & 1.123  & 1.945  \\
 &  & 255 &  4.1' & 2.966  & 5.122  \\
 &  & 285 &  3.7' & 5.459  & 9.405  \\
 &  & 315 &  3.3' & 16.30 & 28.24  \\
 &  & 375 &  2.8' & 49.00 & 85.00 \\
 &  & 435 &  2.4' & 124.0 & 215.0 \\
\hline \hline
\end{tabular}
\end{center}
\caption{The specification of CMB experiment. $\fsky$ is the sky
    coverage, $\theta_{\rm FWHM}$ is the beam width at FWHM,
    $\Delta_\nu^T$ and $\Delta_\nu^P$ represent the sensitivity of each
    channel to the temperature and polarization, respectively.  $\nu$
    indicates the channel frequency of detectors. We assume the
    COrE-like satellite survey, and these parameters can be found in
    \cite{COrE}.}
\label{tb:CMBexp}
\end{table}

For the CMB experiment, we assume the ideal condition in which
foreground removal can be done completely and noise components depend
only on detector noise. If the detector noise is Gaussian white noise,
the noise spectra for the temperature and the polarization fluctuations
are given by
\begin{equation}
    N_\l^{T, P} = \left[ \sum_\nu \left\{
      \left( \theta_{{\rm FWHM},\nu} \Delta_{\nu}^{T, P} \right)^{-2}
      \exp \left( - \frac{\l(\l+1)\theta_{{\rm FWHM},\nu}}{8\ln 2} \right)
      \right\} \right]^{-1} ,  
\end{equation}
where the superscripts $T$ and $P = \{E, B\}$ denote the temperature and
the polarization components, respectively, $\nu$ is the frequency at
each band channel, $\Delta_{\nu}^{T, P}$ is the sensitivity of detector
per pixel at $\nu$ band and $\theta_{{\rm FWHM},\nu}$ represents the
resolution at $\nu$ band.  The values for the COrE-like satellite are
given in Table~\ref{tb:CMBexp}.  On the estimation of the CMB lensing
potential $\psi$, we adopt the noise spectrum of the lensing potential
$N_l^{\psi}$ following the method in \cite{Okamoto:2003} optimally
combining the temperature and the polarization fields. Therefore, the
temperature $T$ and the E-mode polarization in our analysis are unlensed
components and all information on CMB lensing is included in the lensing
potential $\psi$.


\begin{table}[t]
\begin{center}

{\bf The specification of the LSST survey}

\begin{tabular}{cccccc}
\hline \hline
& survey area & sample galaxies & mean redshift & maximum redshift & redshift bin \\
& $f_{\rm sky}$ & $\bar{n}_{\rm g}$ [arcmin$^{-2}$] & $z_{\rm m}$ & $z_{\rm max}$ \\
\hline
LSST & 0.5 & 50  & 1.2 & 3.0 & 5 \\
\hline \hline
\end{tabular}

\vspace{6mm}

{\bf The redshift range for the tomographic survey}

\begin{tabular}{cccccc}
\hline \hline
$i$-th bin & 
1st & 2nd & 3rd & 4th & 5th \\
\hline
redshift 
 & $0 < z \leq \frac{2}{5} z_{\rm m}$ 
 & $\frac{2}{5} z_{\rm m} < z \leq \frac{4}{5} z_{\rm m}$ 
 & $\frac{4}{5} z_{\rm m} < z \leq \frac{6}{5} z_{\rm m}$ 
 & $\frac{6}{5} z_{\rm m} < z \leq \frac{8}{5} z_{\rm m}$ 
 & $\frac{8}{5} z_{\rm m} < z \leq  z_{\rm max} $ \\
\hline \hline
\end{tabular}
\end{center}
\caption{The specification of the LSST survey (Top) and the redshift
   ranges of the $i$-th bin for the case of dividing into five redshift
   bins (Bottom). } \label{tb:survey_params}
\end{table}


For a galaxy clustering measurement ${\rm g}_i$, noise contribution
is associated with the finiteness of the galaxy samples and we assume
that the shot noise is given by
\begin{equation}
    N_\l^{{\rm g}_i} = \delta_{ij} \frac{1}{\bar{n}_i^{\rm A}} ,
\end{equation}
where $\bar{n}_i^{\rm A}$ is the mean surface density of galaxies per
steradian in the $i$-th redshift bin.

For a galaxy weak lensing survey ${\gamma}_i$, we suffer from the 
uncertainties in measuring the shear from galaxy images, one of which
mainly comes from the intrinsic shape ellipticities of galaxies. The
galaxy shapes can be treated statistically and the noise spectrum is
given by
\begin{equation}
    N_\l^{{\gamma}_i} = \delta_{ij} 
        \frac{\sigma_\gamma^2}{n_i^{\rm A}} ,
\end{equation}
where $\sigma_\gamma$ is the uncertainty in the shape measurement and
we adopt the value $\sigma_\gamma = 0.03$ for all redshift bins. 

Here we assume the specification of the Large Synoptic Sky Survey
(LSST\footnote{http://www.lsst.org}) \cite{LSST} and adopt the following
survey parameters: the survey area is 20,000 deg$^2$, the mean redshift
is $z_{\rm m} = 1.2$, the maximum redshift is $z_{\rm max} = 3.0$, and
the number density of sample galaxies is $\ng = 50$
arcmin$^{-2}$. Moreover, we assume a tomographic survey divided into
five redshift bins. The redshift range for each bin is shown in
Table~\ref{tb:survey_params}.
\\

 

\begin{figure}[t]
  \begin{center}
  \includegraphics[clip,width=0.49\textwidth]{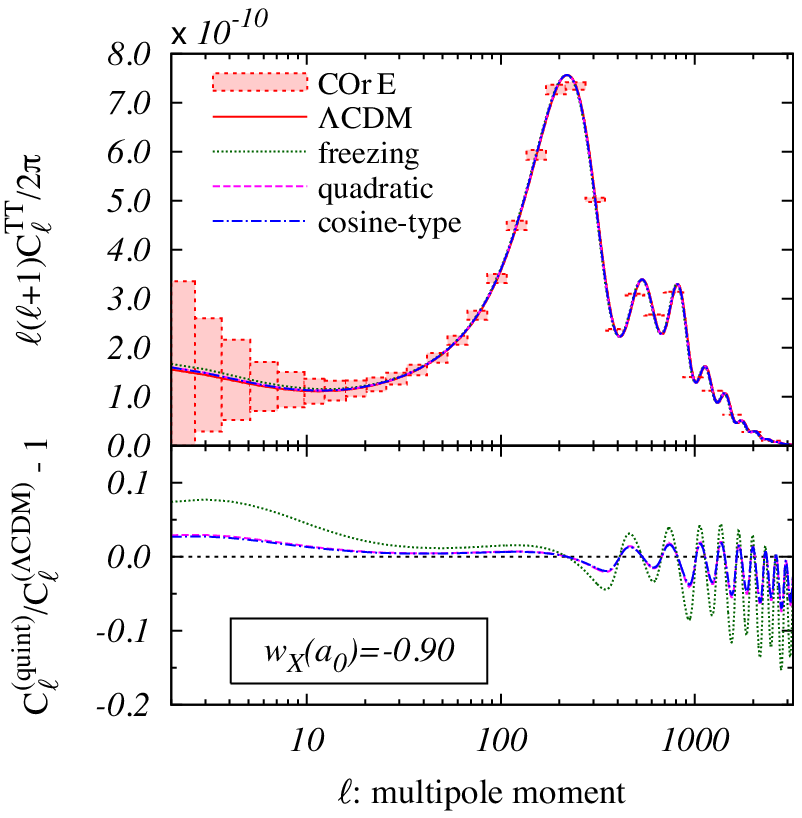}
  \hspace{0.5mm}   
  \includegraphics[clip,width=0.49\textwidth]{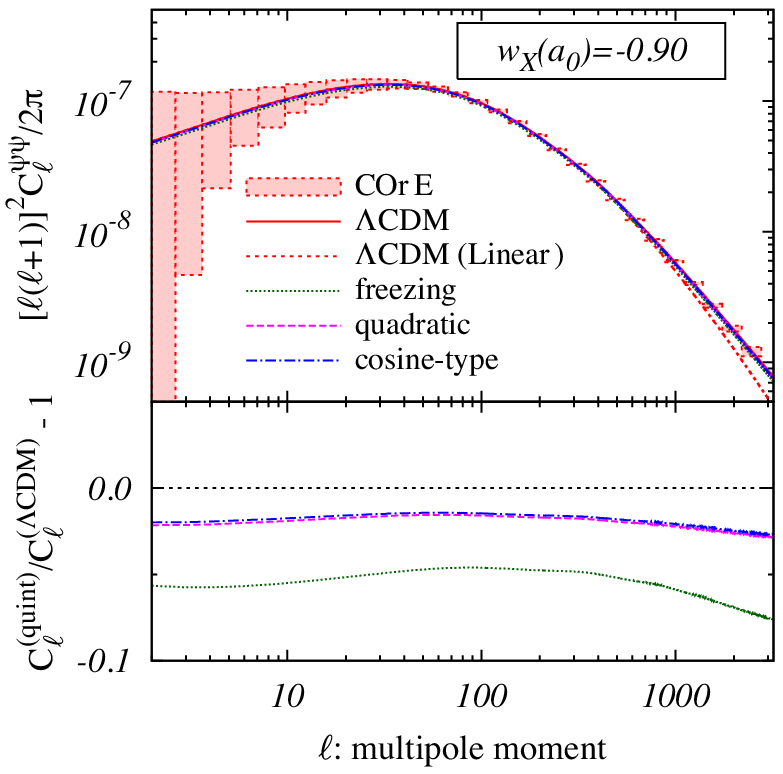}
  \end{center}
  \vspace{-0.5cm}
  \caption{The angular power spectra of CMB for the cases with
  $\Lambda$CDM and some quintessence models (Top). The deviation of the
  power spectra for the quintessence models from that of $\Lambda$CDM
  (Bottom).  The solid (red) line represents the $\Lambda$CDM model, the
  dotted (green) line is for the freezing model, the dashed (magenta)
  line is for the thawing model with the quadratic potential, and the dot-dashed (blue) line is for
  the thawing model with the cosine-type potential. The filled boxes are binning noise spectrum for
  COrE-like CMB experiment. The fiducial models of quintessence are
  chosen such that the present value of the equation of state being
  $w_X(a_0)=-0.90$.  We plot the power spectra for CMB temperature
  (left) and CMB lensing potential (right). For comparison, we also plot
  the linear angular power spectrum of CMB lensing potential for the
  $\Lambda$CDM model with dashed (red) line. } \label{fig:clsCMB}
\end{figure}



\begin{figure}[t]
  \begin{center}
    \resizebox{160mm}{!}{
\includegraphics{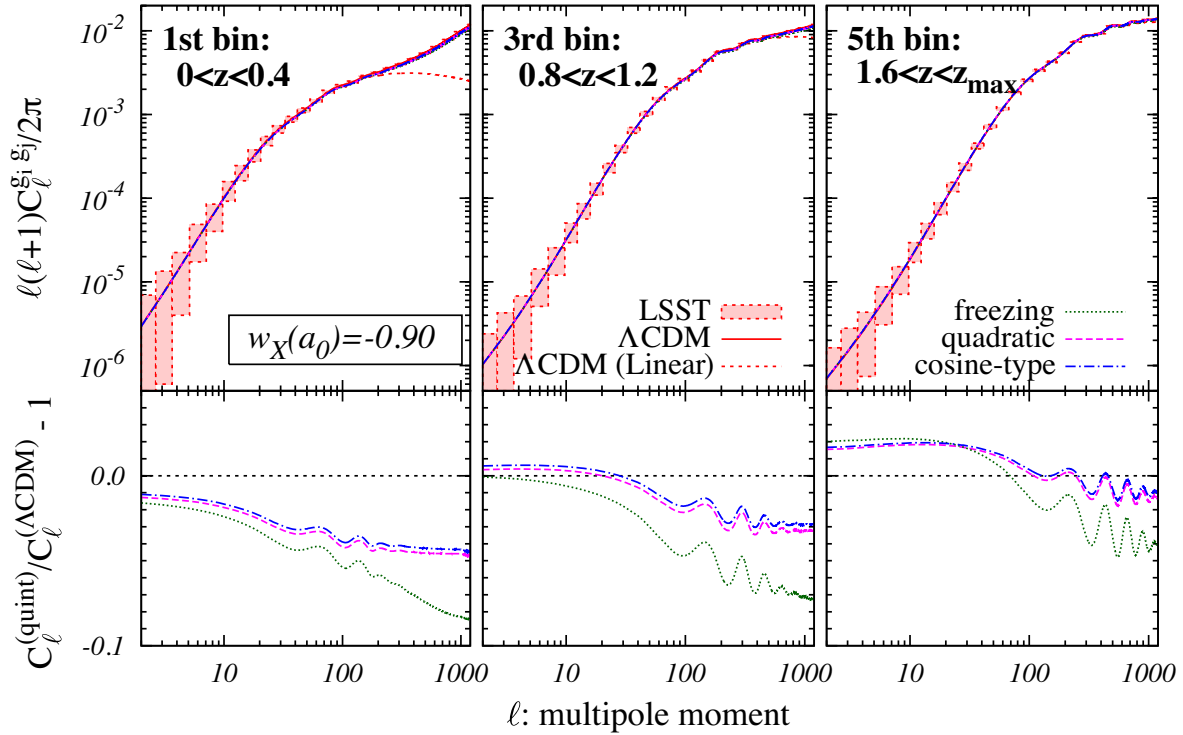}
}
  \end{center}
  \vspace{-0.5cm}
  \caption{ Same as Figure~\ref{fig:clsCMB} except for the angular power
  spectra of the galaxy clustering. The filled boxes are binning noise
  spectrum for the LSST survey. We plot only the auto-correlations ones
  for the 1st, 3rd and 5th bins in the figure. } \label{fig:clsGal}
\end{figure}

\begin{figure}[tbh]
  \begin{center}
    \resizebox{160mm}{!}{
\includegraphics{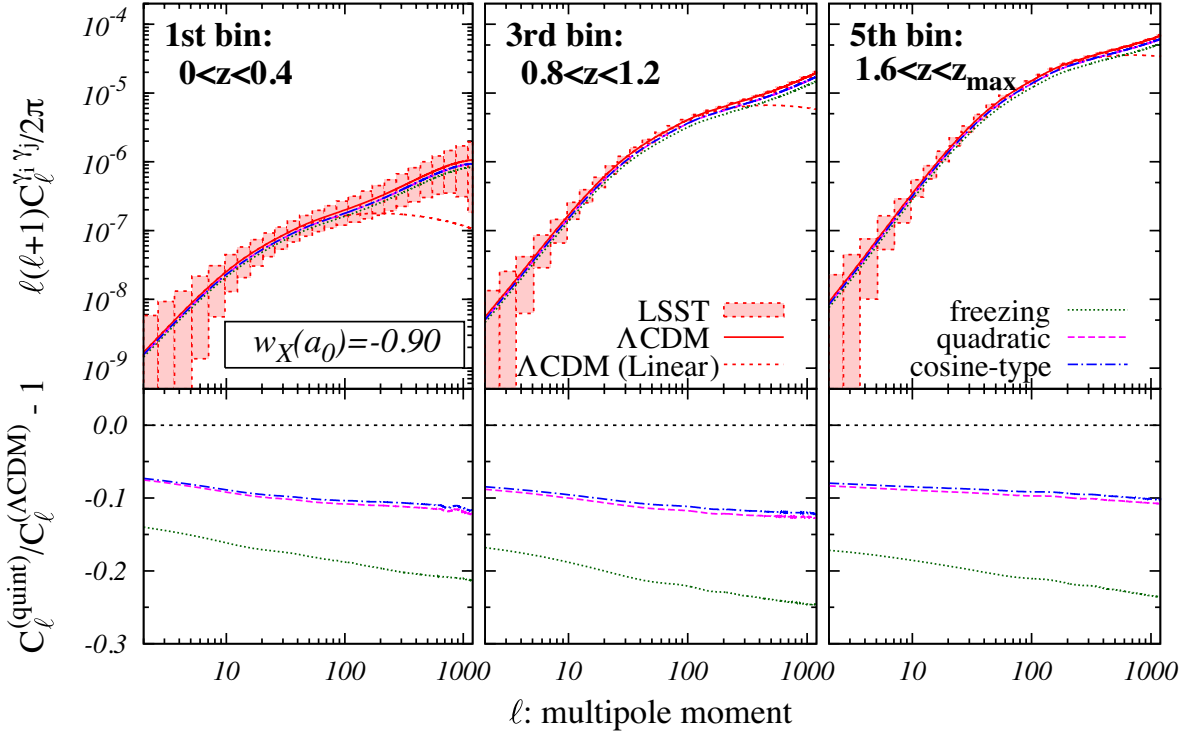}
}
  \end{center}
  \vspace{-0.5cm}
  \caption{Same as Figure~\ref{fig:clsGal} but for the angular power spectra of the galaxy weak lensing
  shear. }
    \label{fig:clsWLen}
\end{figure}

We plot angular power spectra of CMB (temperature mode and lensing
potential), galaxy clustering, and galaxy weak lensing shear in
Figures~\ref{fig:clsCMB}, \ref{fig:clsGal} and \ref{fig:clsWLen},
respectively. Cases for $\Lambda$CDM, freezing and thawing models are
shown and both quadratic and cosine-type potential models
are adopted for thawing models. Model parameters for quintessence are
chosen such that the present value of the equation of state becomes
$w_X=-0.90$.  Other cosmological parameters are taken be those of the
mean values of WMAP7+BAO+$H_0$ analysis for $\Lambda$CDM model
\cite{Komatsu:2010fb}.  In an analogous fashion, we show the spectra
of galaxy clustering and galaxy weak lensing shear in
Figure~\ref{fig:clsGal} and \ref{fig:clsWLen},
respectively. Auto-correlations in the 1st, 3rd and 5th bins are only
depicted for reference.


To discuss the advantage of combining multiple observational data, we
consider the following two cases:

\begin{itemize}
\item{Case I: CMB + SNe}

\hspace{5mm}
$
    {\bm C}_\l^{XY} = \{ 
    C_\l^{TT}, C_\l^{EE}, C_\l^{TE}, C_\l^{\psi\psi}, C_\l^{T\psi}
    \} , \nn 
$
\item{Case II: (CMB $\times$ Galaxy clustering $\times$ Galaxy weak lensing) + SNe}

\hspace{5mm}
$
    {\bm C}_\l^{XY} = \{ 
    C_\l^{TT}, C_\l^{EE}, C_\l^{TE}, C_\l^{\psi\psi}, C_\l^{T\psi}, 
    C_\l^{{\rm g}_i{\rm g}_j}, C_\l^{{\gamma}_i{\gamma}_j}, C_\l^{{\rm g}_i{\gamma}_j},
    C_\l^{T{\rm g}_i}, C_\l^{T{\gamma}_i}, C_\l^{\psi{\rm g}_i}, C_\l^{\psi{\gamma}_i}, 
    \}, \nn
$
\end{itemize}
where Case I corresponds to the constraint from CMB, and Case II
corresponds to that from auto-correlations of CMB, galaxy clustering and
galaxy weak lensing with cross-correlations between different surveys.
Both cases include the information from the SNe survey, for which we
assume JDEM-like specification\footnote{http://science1.nasa.gov/missions/jdem/}.

Then we define the total chi-square of all observables $\chi^2_{\rm tot}$ as
\begin{equation}
        \chi^2_{\rm tot} = \chi^2_{\rm (I\,or\,II)}({\bm C_{\l,\,{\rm mod}}}) 
        + \chi^2_{\rm SNe}({\bm \mu_{i,\,{\rm mod}}}), 
\label{eq:chisq_tot}
\end{equation}
where $\chi^2_{\rm (I\,or\,II)}({\bm C_{\l,\,{\rm mod}}})$ denotes the
chi-square for the angular power spectra given in
Eq.~(\ref{eq:chisq_cls}) for Case\,I or Case\,II, respectively, and
$\chi^2_{\rm SNe}({\bm \mu_{i,\,{\rm mod}}})$ is that from the SNe
survey (given by Eq.~(\ref{eq:chisq_SNe}) in Appendix~\ref{sec:SNe}).


%


We explore the parameter space by Markov-chain Monte-Carlo method and
for this purpose we modify the publicly available code CosmoMC
\cite{Lewis:2002}. To estimate the constraints, we adopt the range of
multipole moments as [$\lmin$,$\lmax$]=[2,2000] for CMB observables
($T$,$E$,$\psi$), and [$\lmin$,$\lmax$]=[2,200] for the LSS observables
(${\rm g}$,$\gamma$), respectively. For the cross-correlations between
CMB and LSS observables, we adopt the same range as that for LSS.

\section{Forecast}
\label{sec:result}
\label{sec:forecast}

Now we discuss the possibility to distinguish the
different quintessence models which give the same present value of $w_X$
from future cosmological surveys.  For this purpose, we derive the constraints
for parameterized dark energy models described in
Section~\ref{sec:perturbEvo} using the mock data generated by assuming
the three fiducial quintessence models.
The first fiducial model is the thawing model in which the potential is
a simple quadratic potential given by Eq.~(\ref{eq:V_phi2}) and labeled
as ``Thawing I.'' The second one is the thawing model with the potential
of a cosine-type given by Eq.~(\ref{eq:V_cosine}) and labeled as
``Thawing II.'' The final one is the freezing model in which the
potential is a typical inverse power-law type given by
Eq.~(\ref{eq:V_tracker}) and labeled as ``Freezing''.  As for the
parameterizations of the dark energy equation of state, we adopt the three
parameterizations given by Eqs.~(\ref{eq:eos1}), (\ref{eq:eos2}), and
(\ref{eq:eos3}).

Since perturbation equations are unstable for dark energy models
crossing $w_X=-1$, we introduce a prior for dark energy equation of
state which forbids the crossing of $w_X=-1$.  We note that this is a
reasonable assumption for this work because quintessence models with
the canonical kinetic term predict its equation of state to be
$w_X \ge -1$. However, to check this prior effect, we compare the models
with or without the prior for ``Parameterization I'' in
Appendix~\ref{sec:const_b} by using only the background quantities.

\begin{figure}[t]
  \begin{center}
    \includegraphics[clip,width=150mm]{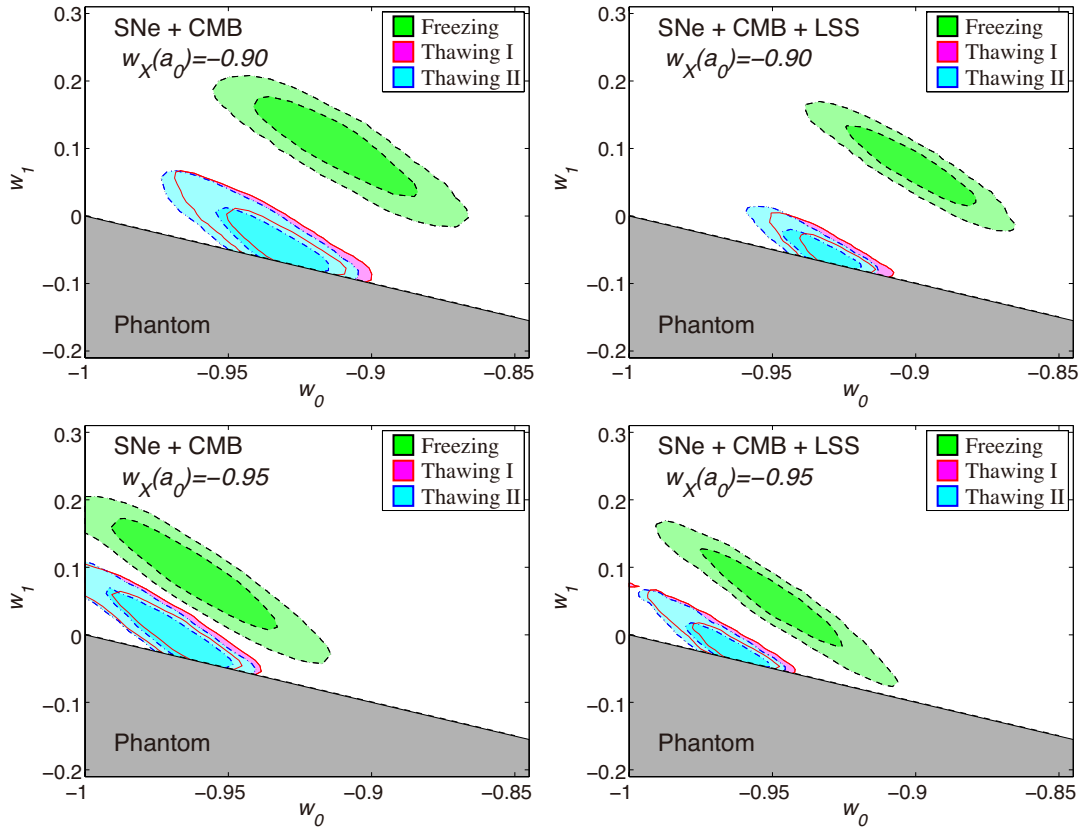}
  \end{center}
   \caption{Expected constraints for Parameterization I from SNe + CMB
   (left) and SNe + CMB + LSS (right). The present value of the equation of
   state for each fiducial model is chosen as $w_X(a_0)=-0.90$ (top) and
   $w_X(a_0)=-0.95$ (bottom), respectively. The contours show the 68\%
   and 95\% confidence level for Freezing (green), Thawing I (magenta),
   and Thawing II (cyan).}  \label{fig:cont_SPL}
\end{figure}

Now we investigate how much we can differentiate the thawing and the
freezing models from future observations of CMB, galaxy clustering and
weak lensing shear. In the analysis, we vary standard six cosmological
parameters plus the model parameters of the dark energy equation of state
$w_X(a)$ for each parameterization and the galaxy bias parameter $b_{\rm
g}(z)$;
\begin{equation}
\{
\Omega_{\rm b}h^2,~
\Omega_{\rm c}h^2,~
\theta_{\rm s},~
\tau^{\rm reion},~
n_{\rm s},~
A_{\rm s}~
\}
+
\left\{
\begin{array}{c}
{\rm model~parameters~of~} w_X(a) 
\end{array}
\right\}
+
\left\{
{\rm galaxy~bias}~b_{\rm g}(z);~ M_{\rm min}
\right\}
\nn
\end{equation}
where $\Omega_{\rm b}$ and $\Omega_{\rm c}$ are the density parameters
of baryon and CDM, $h$ is the dimension-less Hubble parameter,
$\theta_{\rm s}$ is the ratio of the sound horizon to the angular
diameter distance, $\tau^{\rm reion}$ is the optical depth of
reionization, $n_{\rm s}$ and $A_{\rm s}$ are the spectral index and the
amplitude of the initial power spectrum, respectively.  Additionally, we
include the parameter $M_{\rm min}$ shown in Eq.~(\ref{eq:bias}) to
determine the galaxy bias parameter when the observable of galaxy
clustering is included.


\subsection{Difference between Thawing and Freezing models}
\label{sec:diff_pot}

Here we show the constraints on the cosmological parameters and the
parameters for the dark energy equation of state for the three fiducial
models of quintessence, which are the potential models given in
Sec.~\ref{sec:quin} and labeled as Thawing I, Thawing II and Freezing in
the figures hereafter.  In Figure~\ref{fig:cont_SPL}, we adopt
``Parameterization~I'' given in Eq.~(\ref{eq:eos1}) and show the
constraints projected on the $w_0-w_1$ plane.  The fiducial models of
quintessence in the top and bottom panels have different values of
the equation of state at present time with $w_X(a_0)=-0.90$ and
$w_X(a_0)=-0.95$, respectively.  For other cosmological parameters, we
take the mean values from WMAP7+BAO+$H_0$ analysis for the $\Lambda$CDM
model \cite{Komatsu:2010fb} as their fiducial ones.  In the figure, the
gray region corresponds to so-called phantom one with $w_0+w_1<-1$,
which is excluded by the prior mentioned above.  The best-fit values and
derived mean values with marginalized 1-$\sigma$ errors for the
parameters of Parameterization~I are summarized in Table~\ref{tb:spl}.

We can find clear difference between the thawing and the freezing models
on these planes even without the LSS observables. However, for our
purpose to discriminate them, the sign of $w_1$ is crucially important
because $w_1 < 0 $ is expected for the thawing models and $w_1 > 0 $ for
the freezing ones. From this viewpoint, we can conclude that, only for
the case with $w_X(a_0)=-0.90$, they can be discerned at $1\sigma$
confidence level (CL) without the LSS observables. The LSS observables
can further improve the constraint and enables us to discriminate them
at more than $1\sigma$ ($2\sigma$) CL as long as the present equation of
state $w_X(a_0) \gtrsim -0.95~(-0.90)$.

\begin{table}[t]
{ 
\footnotesize
\begin{center}
  \begin{tabular}{lcccccr}
    \hline \hline
    & \multicolumn{2}{c}{$w_0$} 
    & \hspace{3mm}
    & \multicolumn{2}{c}{$w_1$} 
    & \hspace{12mm}
    \\
    \cline{2-3}  
    \cline{5-6}  
    & \makebox[1.2cm]{Best fit} & 
    \multicolumn{1}{c}{68\% limits}
    &
    & \makebox[1.2cm]{Best fit} & 
    \multicolumn{1}{c}{68\% limits}
    & $\chi_{\rm min}^2$
    \\
\hline
\multicolumn{2}{l}{} \\
\multicolumn{2}{l}{\bm{$w_X(a_0)=-0.90$}} \\
\hline
{\bf Freezing} \\
\mbox{\LBCMB} & -0.90 & -0.91  $\pm$  {0.018} & & 0.089 & 0.101  $\pm$  {0.048} & 3.250 \\
\mbox{\LBALL} & -0.90 & -0.90  $\pm$  {0.015} & & 0.084 & 0.074  $\pm$  {0.040} & 6.949 \\
\hline
{\bf Thawing I} \\
\mbox{\LBCMB} & -0.93 & -0.93  $\pm$  {0.013} & & -0.044 & -0.036  $\pm$  {0.035} & 3.445 \\
\mbox{\LBALL} & -0.92 & -0.93  $\pm$  {0.009} & & -0.076 & -0.053  $\pm$  {0.022} & 4.467 \\
\hline
{\bf Thawing II} \\
\mbox{\LBCMB} & -0.92 & -0.94  $\pm$  {0.014} & & -0.075 & -0.031  $\pm$  {0.036} & 3.652 \\
\mbox{\LBALL} & -0.93 & -0.94  $\pm$  {0.009} & & -0.062 & -0.047  $\pm$  {0.023} & 4.921 \\
\hline \hline
\multicolumn{2}{l}{} \\
\multicolumn{2}{l}{\bm{$w_X(a_0)=-0.95$}} \\
\hline 
{\bf Freezing} \\
\mbox{\LBCMB} & -0.96 & -0.96  $\pm$  {0.018} & & 0.079 & 0.086  $\pm$  {0.052} & 3.302 \\
\mbox{\LBALL} & -0.94 & -0.95  $\pm$  {0.017} & & 0.037 & 0.052  $\pm$  {0.048} & 4.404 \\
\hline
{\bf Thawing I} \\
\mbox{\LBCMB} & -0.96 & -0.97  $\pm$  {0.013} & & -0.020 & 0.006   $\pm$  {0.036} & 3.438 \\
\mbox{\LBALL} & -0.96 & -0.97  $\pm$  {0.010} & & -0.041 & -0.013  $\pm$  {0.027} & 4.133 \\
\hline
{\bf Thawing II} \\
\mbox{\LBCMB} & -0.96 & -0.97  $\pm$  {0.013} & & -0.010 & 0.008   $\pm$  {0.036} & 3.410 \\
\mbox{\LBALL} & -0.96 & -0.97  $\pm$  {0.010} & & -0.040 & -0.011  $\pm$  {0.025} & 4.209 \\
\hline \hline
\end{tabular}
\end{center}
}
\caption{ Best-fit values and 68\% confidence limits for
  Parameterization I.  } \label{tb:spl}
\end{table}


%
\begin{figure}[t]
  \begin{center}
\includegraphics[clip,width=150mm]{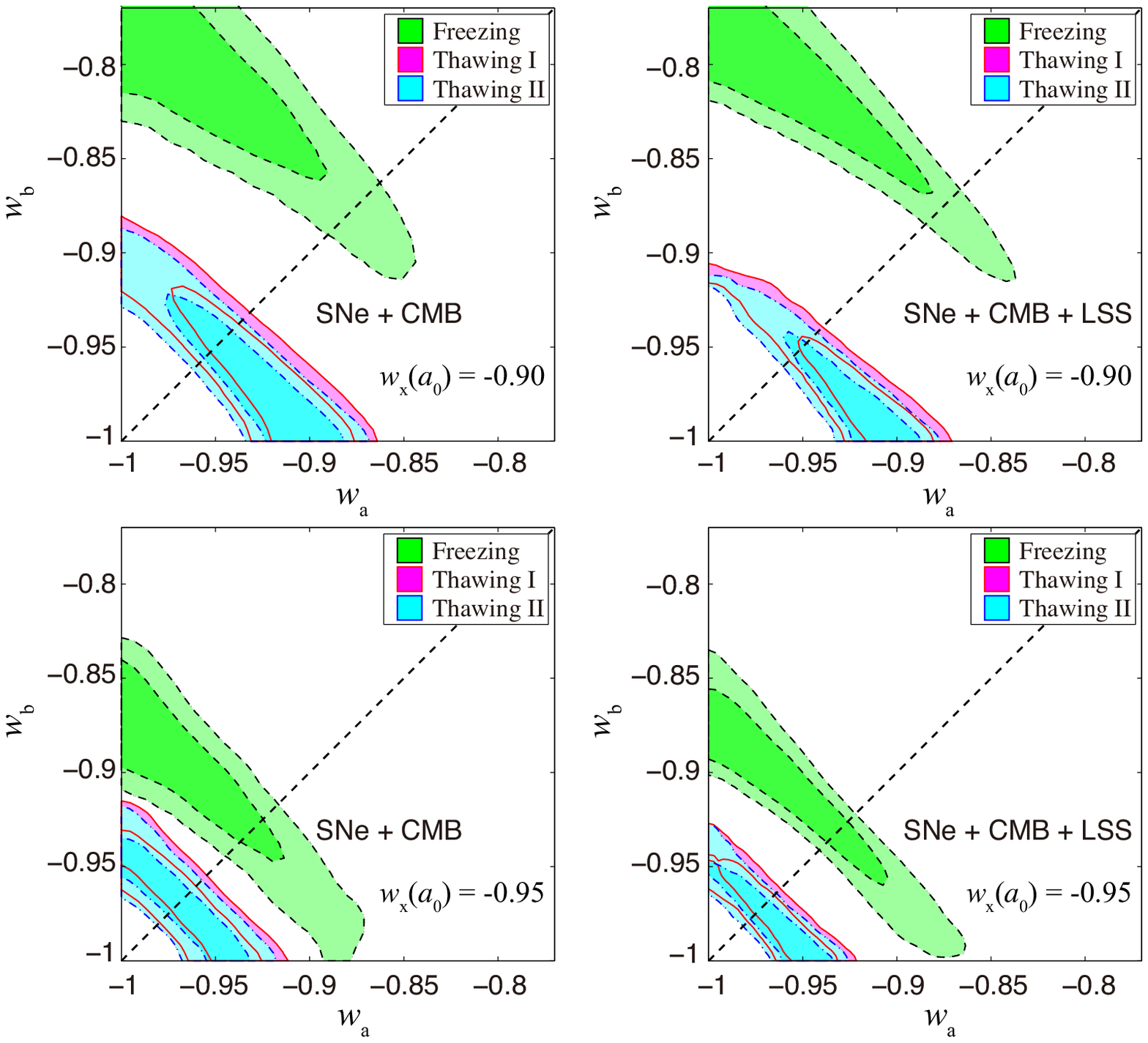}
  \end{center}
  \caption{Same as Figure~\ref{fig:cont_SPL}, except for the constraints
  for Parameterization II with the value of $q=0.8$.}
 \label{fig:cont_HM0.8}
\end{figure}
\begin{figure}[t]
  \begin{center}
\includegraphics[clip,width=150mm]{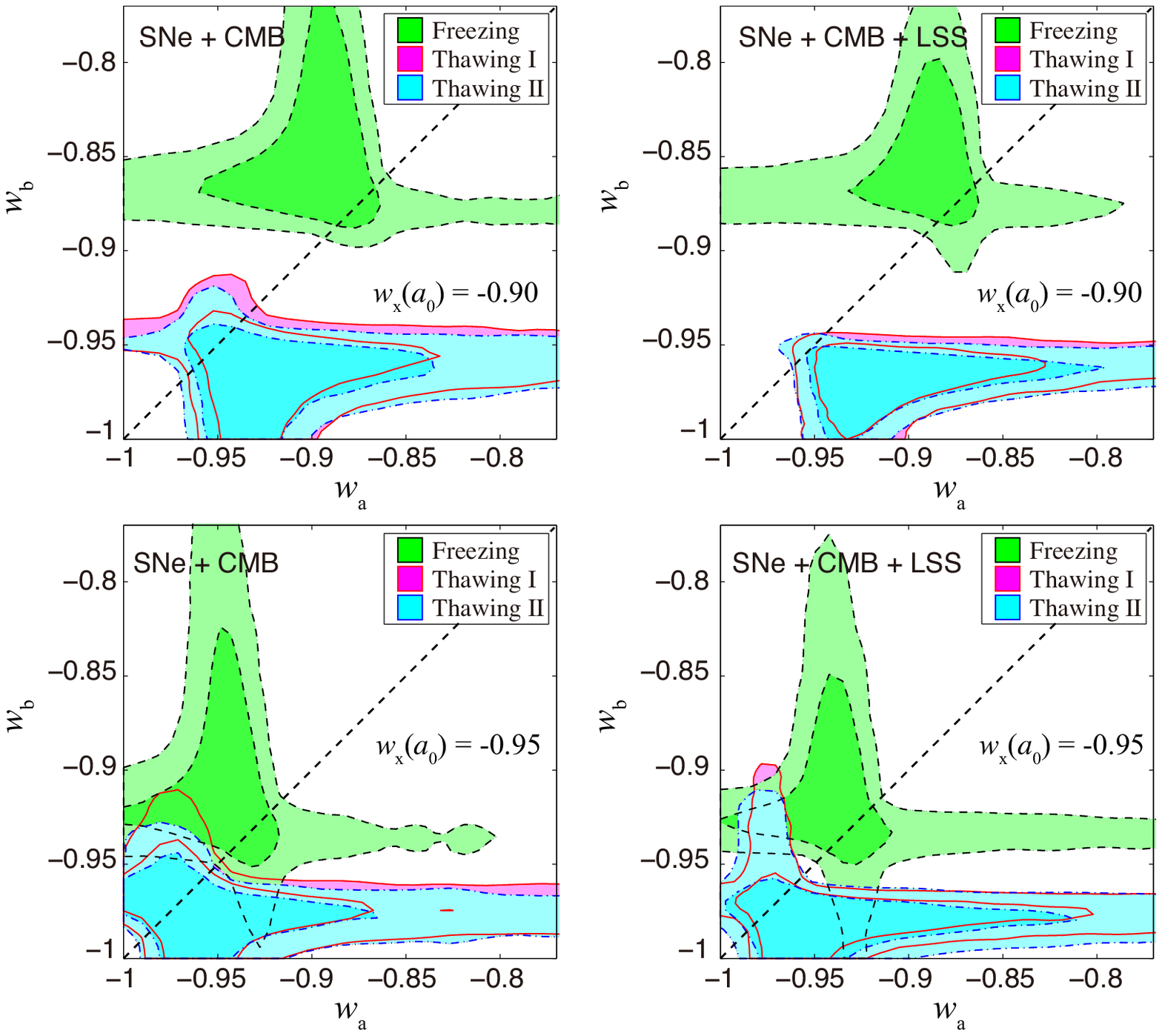}
  \end{center}
  \caption{Same as Figure~\ref{fig:cont_SPL}, except for the constraints
  for Parameterization II with the value of $q=100$.}
 \label{fig:cont_HM100}
\end{figure}

\begin{table}[t]
{ 
\footnotesize
\begin{center}
  \begin{tabular}{lccccccccr}
    \hline \hline
    & \multicolumn{2}{c}{$w_a$} 
    & \hspace{1mm}
    & \multicolumn{2}{c}{$w_b$} 
    & \hspace{1mm}
    & \multicolumn{2}{c}{$a_s$} 
    & \hspace{12mm}
    \\
    \cline{2-3}  
    \cline{5-6}  
    \cline{8-9}  
    & \makebox[1.0cm]{Best fit} & 
    \multicolumn{1}{c}{68\% limits}
    &
    & \makebox[1.0cm]{Best fit} & 
    \multicolumn{1}{c}{68\% limits}
    &
    & \makebox[1.0cm]{Best fit} & 
    \multicolumn{1}{c}{68\% limits}
    & $\chi_{\rm min}^2$
    \\
\hline
\multicolumn{2}{l}{} \\ 
\multicolumn{2}{l}{\bm{$w_X(a_0)=-0.90~(q=0.8)$}} \\
\hline
{\bf Freezing} \\
\mbox{\LBCMB} & -0.99 & -0.94  $\pm$  {0.039} & & -0.75 & -0.81  $\pm$  {0.036} & & 0.38 & 0.60  $\pm$  {0.19} & 3.479 \\
\mbox{\LBALL} & -0.98 & -0.93  $\pm$  {0.038} & & -0.77 & -0.83  $\pm$  {0.028} & & 0.42 & 0.63  $\pm$  {0.18} & 6.322 \\
\hline
{\bf Thawing I} \\
\mbox{\LBCMB} & -0.91 & -0.93  $\pm$  {0.029} & & -0.98 & -0.95  $\pm$  {0.029} & & 0.67 & 0.67  $\pm$  {0.19} & 3.619 \\
\mbox{\LBALL} & -0.91 & -0.92  $\pm$  {0.024} & & -0.99 & -0.97  $\pm$  {0.022} & & 0.55 & 0.69  $\pm$  {0.19} & 6.056 \\
\hline
{\bf Thawing II} \\
\mbox{\LBCMB} & -0.91 & -0.94  $\pm$  {0.028} & & -0.99 & -0.96  $\pm$  {0.027} & & 0.55 & 0.68  $\pm$  {0.19} & 3.710 \\
\mbox{\LBALL} & -0.90 & -0.93  $\pm$  {0.024} & & -0.99 & -0.97  $\pm$  {0.022} & & 0.88 & 0.70  $\pm$  {0.19} & 6.003 \\
\hline \hline
\multicolumn{2}{l}{} \\ 
\multicolumn{2}{l}{\bm{$w_X(a_0)=-0.95~(q=0.8)$}} \\
\hline
{\bf Freezing} \\
\mbox{\LBCMB} & -0.99 & -0.96  $\pm$  {0.032} & & -0.87 & -0.90  $\pm$  {0.034} & & 0.57 & 0.62  $\pm$  {0.19} & 3.973 \\
\mbox{\LBALL} & -0.99 & -0.95  $\pm$  {0.033} & & -0.86 & -0.91  $\pm$  {0.034} & & 0.47 & 0.64  $\pm$  {0.19} & 4.110 \\
\hline
{\bf Thawing I} \\
\mbox{\LBCMB} & -0.95 & -0.97  $\pm$  {0.020} & & -0.98 & -0.96  $\pm$  {0.020} & & 0.93 & 0.66  $\pm$  {0.19} & 3.284 \\
\mbox{\LBALL} & -0.98 & -0.97  $\pm$  {0.018} & & -0.96 & -0.97  $\pm$  {0.018} & & 0.85 & 0.66  $\pm$  {0.19} & 4.772 \\
\hline
{\bf Thawing II} \\
\mbox{\LBCMB} & -0.97 & -0.97  $\pm$  {0.019} & & -0.96 & -0.96  $\pm$  {0.019} & & 0.70 & 0.66  $\pm$  {0.19} & 3.357 \\
\mbox{\LBALL} & -0.96 & -0.97  $\pm$  {0.017} & & -0.98 & -0.97  $\pm$  {0.017} & & 0.64 & 0.67  $\pm$  {0.19} & 4.700 \\
\hline \hline
\multicolumn{2}{l}{} \\ 
\multicolumn{2}{l}{$\bm{w_X(a_0)=-0.90~(q=100)}$} \\
\hline
{\bf Freezing} \\
\mbox{\LBCMB} & -0.89 & -0.90  $\pm$  {0.039} & & -0.85 & -0.84  $\pm$  {0.039} & & 0.66 & 0.64  $\pm$  {0.21} & 2.874 \\
\mbox{\LBALL} & -0.89 & -0.89  $\pm$  {0.029} & & -0.85 & -0.84  $\pm$  {0.026} & & 0.57 & 0.64  $\pm$  {0.20} & 6.150 \\
\hline
{\bf Thawing I} \\
\mbox{\LBCMB} & -0.92 & -0.90  $\pm$  {0.056} & & -0.96 & -0.96  $\pm$  {0.017} & & 0.73 & 0.74  $\pm$  {0.19} & 3.029 \\
\mbox{\LBALL} & -0.90 & -0.90  $\pm$  {0.060} & & -0.96 & -0.97  $\pm$  {0.012} & & 0.77 & 0.79  $\pm$  {0.12} & 3.868 \\
\hline
{\bf Thawing II} \\
\mbox{\LBCMB} & -0.91 & -0.89  $\pm$  {0.065} & & -0.96 & -0.96  $\pm$  {0.016} & & 0.82 & 0.76  $\pm$  {0.19} & 3.111 \\
\mbox{\LBALL} & -0.90 & -0.88  $\pm$  {0.063} & & -0.97 & -0.97  $\pm$  {0.011} & & 0.80 & 0.79  $\pm$  {0.14} & 3.885 \\
\hline \hline
\multicolumn{2}{l}{} \\ 
\multicolumn{2}{l}{$\bm{w_X(a_0)=-0.95~(q=100)}$} \\
\hline
{\bf Freezing} \\
\mbox{\LBCMB} & -0.94 & -0.94  $\pm$  {0.036} & & -0.88 & -0.90  $\pm$  {0.046} & & 0.44 & 0.61  $\pm$  {0.21} & 3.025 \\
\mbox{\LBALL} & -0.94 & -0.93  $\pm$  {0.042} & & -0.91 & -0.91  $\pm$  {0.034} & & 0.56 & 0.63  $\pm$  {0.22} & 4.114 \\
\hline
{\bf Thawing I} \\
\mbox{\LBCMB} & -0.87 & -0.92  $\pm$  {0.072} & & -0.97 & -0.97  $\pm$  {0.016} & & 0.95 & 0.76  $\pm$  {0.21} & 3.090 \\
\mbox{\LBALL} & -0.94 & -0.91  $\pm$  {0.072} & & -0.98 & -0.97  $\pm$  {0.025} & & 0.86 & 0.75  $\pm$  {0.20} & 3.914 \\
\hline
{\bf Thawing II} \\
\mbox{\LBCMB} & -0.91 & -0.92  $\pm$  {0.072} & & -0.97 & -0.97  $\pm$  {0.013} & & 0.94 & 0.77  $\pm$  {0.20} & 3.126 \\
\mbox{\LBALL} & -0.95 & -0.91  $\pm$  {0.071} & & -0.98 & -0.98  $\pm$  {0.012} & & 0.86 & 0.79  $\pm$  {0.19} & 3.852 \\
\hline \hline
\end{tabular}
\end{center}
}
\caption{Best-fit values and 68\% confidence limits for
    Parameterization II.  } \label{tb:hm}
\end{table}


We also study constraints adopting another parameterization for the equation of
state to see its dependence on the parameterization.  In
Figure~\ref{fig:cont_HM0.8} and \ref{fig:cont_HM100}, we show the
results for ``Parameterization II'' given by Eq.~(\ref{eq:eos2}).
Although this parameterization contains four parameters $w_a$, $w_b$,
$a_s$ and $q$, we fix the value of $q$ as $0.8$ and $100$ for
Figure~\ref{fig:cont_HM0.8} and \ref{fig:cont_HM100}, respectively. The
different values of $q$ give the different slope or interval of the
evolution for the dark energy equation of state. Small $q$ gives a mild and
long-term shift of $w_X(a)$ while large $q$ gives a rapid and short-term
shift. The figures represent constraints on the $w_a-w_b$ plane. For the
same reason as discussed above, we prohibit the phantom region of dark
energy equation of state with $w_X(a)<-1$, which corresponds to $w_a<-1$
and $w_b<-1$ in this parameterization. 
The best-fit values and derived mean values with marginalized 1$\sigma$ errors for 
the parameters of Parameterization~II are summarized in Table~\ref{tb:hm}.

In the case of $q=0.8$, the thawing and the freezing models show up in
different parameter regions even without the LSS observables. Note,
however, that the allowed region should appear in the down-right side
($w_a > w_b$) on the $w_a-w_b$ plane for thawing models while in the
up-left side ($w_b > w_a$) for freezing ones. In this respect, we
conclude from Figure~\ref{fig:cont_HM0.8} that we cannot distinguish
them without the LSS observables.  Only when combing the LSS
observables, they can be differentiated at $1\sigma$ CL only for the
case with $w_X(a_0)=-0.90$.  For the case with $w_X(a_0)=-0.95$, we
cannot make a definite statement for the difference between the thawing
and the freezing models.

Let us move on to the case of $q=100$. Apparently, the distributions of
allowed regions drastically change (Figure~\ref{fig:cont_HM100}). This
is because the evolution of the equation of state shows an instantaneous
transition and the constraints are almost determined by the value of
equation of state in the fiducial model at the present epoch. If the
transition of the equation of state occurs early enough, it is almost
constant, $w_X(a)\simeq w_a$, over the observationally relevant epoch.
However, the results are almost the same as those for the case of
$q=0.8$ in terms of discrimination of the thawing and the freezing
models.  As shown in Figure~\ref{fig:cont_HM100}, they can be
distinguished at $1\sigma$ CL only for the case with $w_X(a_0)=-0.90$
while they are indistinguishable for the case with $w_X(a_0)=-0.95$.

To summarize, as for parameterization II, there is no substantial
difference between the cases with $q = 0.8$ and $q = 100$ in order to
distinguish between the thawing and the freezing models on the $w_a-w_b$
plane.

\clearpage

\begin{figure}[t]
  \begin{center}
\includegraphics[clip,width=150mm]{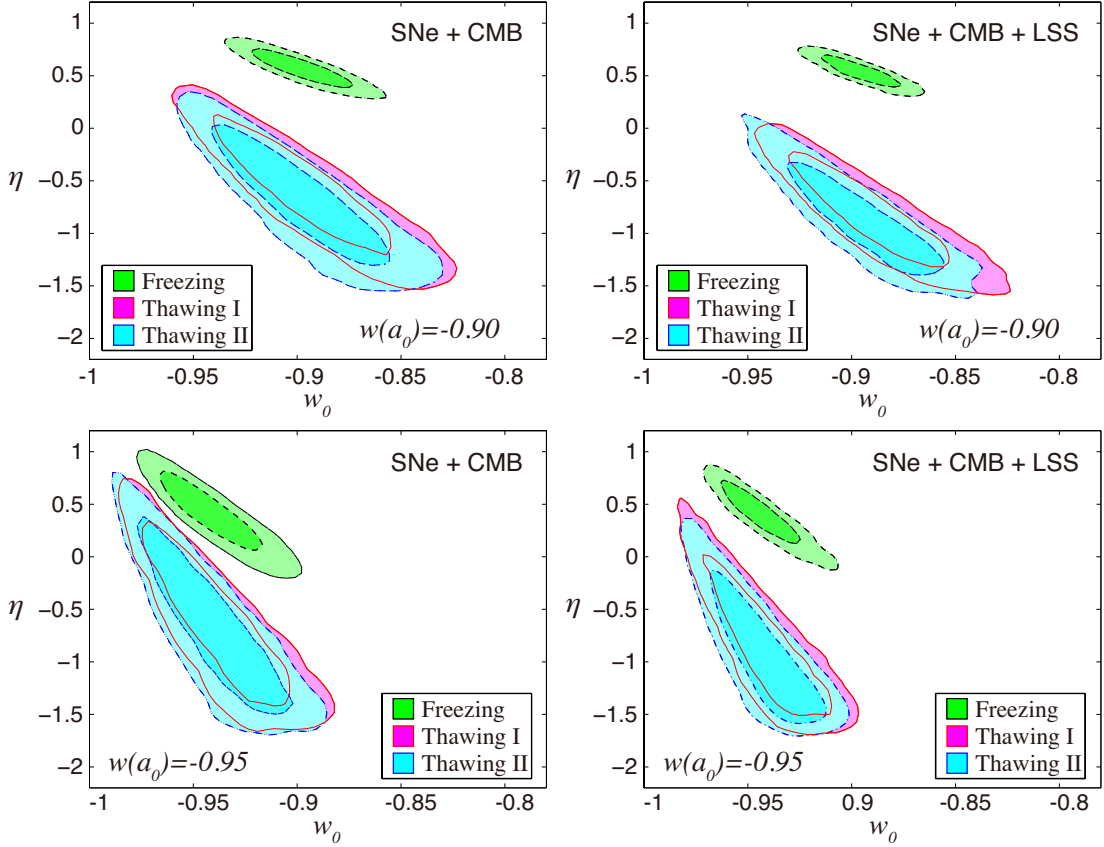}
  \end{center}
  \caption{Same as Figure~\ref{fig:cont_SPL}, but for the constraints for Parameterization III.}
 \label{fig:cont_slow}
\end{figure}

\subsection{The sign of the effective mass squared within thawing models}
\label{sec:thaw_pot}

In this subsection, we investigate the possibility of distinguishing the
different thawing quintessence models, i.e., quadratic and cosine-type
potentials. For this purpose, we focus on the sign of the effective mass
squared of the potentials and adopt Parameterization III given in
Eq.~(\ref{eq:eos3}).

In Figure~\ref{fig:cont_slow}, we show the results for
Parameterization~III. While the main purpose of this analysis is to
investigate whether we can differentiate the sign of the effective mass
of the potential for the thawing model, we also show the constraint on
the freezing model in the same figures for comparison.  The best-fit
values and derived mean values with marginalized 1$\sigma$ errors for
the parameters of Parameterization~III are summarized in
Table~\ref{tb:slow}.

Despite our expectation, constraints on the parameter $\eta$ are
indistinguishable within the thawing models, i.e., quadratic and cosine-type 
potentials. This is because the parameter $\eta$ not only serves as potential
curvature but also affects the time evolution of the equation of state. In
other words, the information about $\eta$ is obtained mainly through
Eq.~(\ref{eq:dwdlna}) and not through Eq.~(\ref{eq:cs2_eta}). Even
if we combine the LSS observables, we cannot see the difference of the
effective mass squared between the two thawing models.

\begin{table}[t]
{ 
\footnotesize
\begin{center}
  \begin{tabular}{lcccccr}
    \hline \hline
    & \multicolumn{2}{c}{$w_0$} 
    & \hspace{4mm}
    & \multicolumn{2}{c}{$\eta$} 
    & \hspace{12mm}
    \\
    \cline{2-3}  
    \cline{5-6}  
    & \makebox[1.2cm]{Best fit} & 
    \multicolumn{1}{c}{68\% limits}
    &
    & \makebox[1.2cm]{Best fit} & 
    \multicolumn{1}{c}{68\% limits}
    & $\chi_{\rm min}^2$ 
    \\
    \hline
\multicolumn{2}{l}{} \\ 
\multicolumn{2}{l}{\bm{$w_X(a_0)=-0.90$}} \\
\hline 
{\bf Freezing} \\
\mbox{\LBCMB} & -0.90 & -0.90  $\pm$  {0.015} & & -0.57 & -0.55  $\pm$  {0.12} & 2.902 \\
\mbox{\LBALL} & -0.90 & -0.90  $\pm$  {0.012} & & -0.59 & -0.54  $\pm$  {0.10} & 7.108 \\
\hline
{\bf Thawing I} \\
\mbox{\LBCMB} & -0.91 & -0.89  $\pm$  {0.028} & & 0.39 & 0.60  $\pm$  {0.41} & 3.152 \\
\mbox{\LBALL} & -0.90 & -0.89  $\pm$  {0.025} & & 0.70 & 0.83  $\pm$  {0.34} & 4.257 \\
\hline
{\bf Thawing II} \\
\mbox{\LBCMB} & -0.91 & -0.90  $\pm$  {0.027} & & 0.40 & 0.68  $\pm$  {0.41} & 3.202 \\
\mbox{\LBALL} & -0.88 & -0.90  $\pm$  {0.025} & & 1.11 & 1.06  $\pm$  {0.34} & 4.570 \\
\hline \hline
\multicolumn{2}{l}{} \\ 
\multicolumn{2}{l}{\bm{$w_X(a_0)=-0.95$}} \\
\hline
{\bf Freezing} \\
\mbox{\LBCMB} & -0.95 & -0.94  $\pm$  {0.016} & & -0.57 & -0.41  $\pm$  {0.24} & 3.241 \\
\mbox{\LBALL} & -0.95 & -0.94  $\pm$  {0.013} & & -0.39 & -0.40  $\pm$  {0.20} & 4.646 \\
\hline
{\bf Thawing I} \\
\mbox{\LBCMB} & -0.95 & -0.94  $\pm$  {0.022} & & 0.37 & 0.63  $\pm$  {0.52} & 3.284 \\
\mbox{\LBALL} & -0.95 & -0.94  $\pm$  {0.019} & & 0.71 & 0.84  $\pm$  {0.48} & 4.116 \\
\hline
{\bf Thawing II} \\
\mbox{\LBCMB} & -0.95 & -0.95  $\pm$  {0.029} & & 0.42 & 0.30  $\pm$  {0.99} & 3.169 \\
\mbox{\LBALL} & -0.93 & -0.94  $\pm$  {0.017} & & 1.08 & 0.92  $\pm$  {0.46} & 4.017 \\
\hline \hline 
\end{tabular}
\end{center}
}
\caption{Best-fit values and 68\% confidence limits for
  Parameterization~III.}  \label{tb:slow}
\end{table}

\subsection{Discussion}
\label{sec:discuss}

We performed analyses assuming some parameterizations for the dark
energy equation of state.  Constraints on each parameter plane depend on
the parameterization, and the degree of how we can distinguish the
different potential models also varies by the parameterizations. Here we
discuss what parameterization of the dark energy equation of state we should
adopt in order to discriminate quintessence models.

As shown in the previous section, Parameterization I enables us to
discriminate them at more than $1\sigma$ ($2\sigma$) CL for the present
equation of state $w_X(a_0) \gtrsim -0.95~(-0.90)$. On the other hand,
in terms of Parameterization II with $q = 0.8$ or $q = 100$, they can be
distinguished at $1\sigma$ CL only for the case with $w_X(a_0)=-0.90$
while they are indistinguishable for the case with $w_X(a_0)=-0.95$.  In
the analysis, we fixed the parameter $q$ (albeit varied $a_s$) simply
because we need enormous numerical efforts due to the degeneracies in
the parameters $q$ and $a_s$.  However, we would expect that, if we vary
both $q$ and $a_s$ simultaneously, the degeneracy between these
parameters would in turn give another degeneracy in other parameters
such as $w_a$ and $w_b$ and loosen the constraints. Therefore, for the
purpose of distinguishing the thawing and the freezing models, our
analysis suggests that Parameterization~I would be more suitable
compared to Parameterization~II.

For the different potentials within the thawing model, we cannot see the
apparent distinction between them from the results with the
parameterizations considered in this paper. The equations of state
of these two thawing models predict almost the same time evolution,
however they have enormously different feature in the parameter $\eta$
given in Eq.~(\ref{eq:eta}). The parameter $\eta$ corresponds to the
curvature of the effective mass squared of potential and those of
Thawing I and Thawing II have opposite signs. To focus on these aspects,
we have adopted Parameterization III and have expected that the
parameter $\eta$ of this parameterization could reflect the
difference between these two models effectively.  However, contrary to
the fact that Thawing I has a positive sign and Thawing II has a negative 
sign of $\eta$, the difference appears only between thawing ($\eta < 0$)
and freezing ($\eta > 0$).  Absolutely, the effects on the density
perturbations due to the differences of the potential's curvature are
reflected through Eq.~(\ref{eq:cs2_eta}), but the constraints seem to
reflect the evolution of the equation of state only. Parameterization
III cannot necessarily well capture the differences of curvature and we
may have to consider another parameterization which can trace the
property of potential more efficiently to distinguish the different
potentials within the thawing models.\footnote{In
Ref.~\cite{Chiba:2013}, they perform the analyses to put the bounds on
models of quintessence from current data by using the parameterization
of the dark energy equation of state. The parameterization also includes
parameters which can reflect the curvature of the potential and which
controls the evolution of the equation of state, separately
(e.g., \cite{Dutta:2008,Chiba:2009}), and an analysis with such a
parameterization might be a breakthrough for the classification within
the thawing models, though the parameterization is a bit
complicated in a present form.}

\bigskip Now several comments on the benefits from large-scale structure
surveys are in order.  The different potentials of quintessence models
modify the evolution of density fluctuations besides the expansion rate
of the Universe. Although the temperature fluctuations of CMB can be
affected by the density fluctuations of dark energy at late times
through the late-ISW effect, we cannot expect large signal-to-noise
ratio from this effect due to the cosmic variance.  On the other hand,
observables from large-scale structure surveys such as galaxy clustering
and galaxy weak lensing directly trace the evolution at low redshifts
where dark energy becomes the dominant component of the Universe, thus
in general, we can extract the information about the evolution of the
density fluctuations more effectively with LSS surveys. Furthermore, the
observable from the SNe survey provides only the information of
background quantities through distance measurement. Actually the
information from the SNe survey plays the most dominant role in the
constraint of the equation of state parameters, but the information
about density fluctuations is independent and helps to break the
degeneracies among model parameters. Additionally, the different
potentials of the thawing models show a similar evolution of the equation of
state and the difference between them appears in the evolution of
density fluctuations. Therefore, the information about the density
fluctuations is absolutely necessary to distinguish the potential models
more finely.

In fact, although the inclusion of LSS data has not drastically
improved the constraints on dark energy parameters in our analysis, such
improvement is useful to discriminate the thawing and the freezing
models, as discussed in the subsection \ref{sec:diff_pot}. For example,
in the case of Parameterization I, we can differentiate the models at
$1\sigma$ confidence level (CL) without the LSS observables only for the
case with $w_X(a_0)=-0.90$.  However, the inclusion of the LSS
observables enables us to discriminate them at more than $1\sigma$
($2\sigma$) CL as long as the present equation of state $w_X(a_0)
\gtrsim -0.95~(-0.90)$. We should also note that there is a possibility
of putting tighter constraints on dark energy from future survey of LSS
such as Euclid \cite{Euclid}\footnote{http://sci.esa.int/euclid}. 
Euclid is a satellite telescope and provides spectroscopic redshift
information. The spectroscopic redshift survey provides more information
than photometric redshift surveys and allows us to measure the
three-dimensional galaxy power spectra. Therefore, the spectroscopic
survey by Euclid will have a potential to put tighter constraints on
dark energy, which will be a next step of this work.

On the other hand, constraints from the LSS observables can be affected
by the designs of survey, for example observing redshift range, the
number or width of redshift bins, systematics of surveys, and so on.
One of the most serious theoretical uncertainties of LSS survey is the
non-linearity of matter power spectra on small scales. In such scales,
we have to take care of mode-couplings for the estimation of covariance
matrix and the misestimation may lead to an incorrect
constraint. However we used the information only in relatively large
scale region where the linear theory is reasonably satisfied for the
observables of the galaxy clustering and galaxy weak lensing. Therefore, 
the systematics due to the non-linearity should be small in our
analysis, and the covariance matrix used in this paper would give a
reasonable estimate. In addition to the above uncertainties, there are
various systematics for galaxy weak lensing shear survey, for example,
on the measurement of galaxy shear and the uncertainties due to
photometric redshift measurement. They make crucial systematics for
constraints on dark energy and/or other parameters
\cite{Huterer:06,Das:2011}. In this paper, we take into account those
systematics as the uncertainties of the observables.  Therefore, the
effects from the other systematics should not affect our results much.

\section{Summary}
\label{sec:summary}

We performed the analysis to investigate how we can distinguish the
different models of quintessence from cosmological surveys in a next few
decades. In this paper we adopted three types of quintessence potentials
as the fiducial models and explored the cosmological parameter space
with MCMC method for each fiducial model. For fitting models, we assumed
some parameterizations for the dark energy equation of state, and considered
future cosmological surveys such as those of the type-Ia supernovae
(JDEM), CMB (COrE) and large scale structures (LSST).

Regarding the differentiations of the thawing and the freezing
quintessence models, we can discriminate them at more than $1\sigma$
($2\sigma$) CL for the case with the present equation of state $w_X(a_0)
\gtrsim -0.95~ (-0.90)$, when we make use of Parameterization I. On the
other hand, weaker constraint is obtained for Parameterization II for
fixed $q$'s. They can be distinguished at $1\sigma$ CL only for the case
with $w_X(a_0)=-0.90$ while they are indistinguishable for the case with
$w_X(a_0)=-0.95$. Thus, we conclude that Parameterization~I would be
more suitable, compared Parameterization~II, in order to distinguish the
thawing and the freezing quintessence models.

For further discrimination of thawing quintessence models, we considered
quadratic and cosine-type potentials. In the analysis using
parameterization I and II, we could not see clear differences between
them. Then we considered the parameterization III which is focused on
the curvature or the sign of the effective mass squared of the
potential. Unfortunately, again we could not see clear differences even
with this parameterization.  Contrary to our expectation, it is found
that this parameterization can not reflect the curvature of potentials
effectively and the sign of curvature is not imprinted into the model
parameters. Therefore, we conclude that we have to invent a new way of
parameterization which can express the difference of potential's
curvature more effectively to distinguish the potentials of thawing
models.

In this work, we concentrated on constraints on dark energy but we can
apply the method here to other subjects, for example the test of
gravity, neutrinos, warm dark matter models, and so on, to compare the
constraints on different models.  Moreover the spectroscopic surveys of
large scale structure will provide more information and put tighter
constraints. However, distinguishing quintessence potentials within the
thawing models will not be improved by including spectroscopic surveys
because a crucial problem seems to be in the parameterization of the
equation of state, which will be a subject of a future work.


\bigskip\bigskip
\section*{Acknowledgements}
We wish to thank S. Yokoyama and T. Chiba for discussion and comments. 
We acknowledge support from the JSPS Grant-in-Aid for Fellows (YT); the
JSPS Grant-in-Aid for Scientific Research under Grant Nos.~24340048
(KI), 23740195 (TT), and 25287054 (MY); the JSPS Grant-in-Aid for
Scientific Research on Innovative Areas No.~24111706 (MY); Grant-in-Aid
for Nagoya University Global COE Program ``Quest for Fundamental
Principles in the Universe: from Particles to the Solar System and the
Cosmos'', Kobayashi-Maskawa Institute for the Origin of Particles and
the Universe; Nagoya University for providing computing resources useful
in conducting the research reported in this paper, Grant-in-Aid for
Scientific Research.

\clearpage

\appendix

\bigskip\bigskip\bigskip
\noindent
{\Large \bf Appendix}

\section{Analysis with background quantities}
\label{sec:background}

Here we present future constraints for the quintessence models discussed
in the main text only by using the information on the background
evolution. One of the merits to perform such an analysis lies in the
fact that we can study the effects of a prior forbidding the
phantom-crossing region. To obtain the constraints from background
quantities, we use projected data from COrE \cite{COrE} for CMB, and
JDEM \cite{DETF}
for SNe, and bigBOSS \cite{Schlegel:2011zz} for BAO,
respectively. 

\subsection{CMB}

For the purpose to estimate the confidence limit of each parameter only
from background quantities, it is useful to introduce some parameters
characterizing the CMB spectra following the method in
\cite{Mukherjee:2008}. In this way we can evaluate the confidence limit
of each parameter without calculating the angular power spectra of CMB
directly.

Here, we use two parameters effectively describing the information
contained in the CMB spectrum, which are proposed in
\cite{Mukherjee:2008}:
\begin{eqnarray}
R &\equiv& \sqrt{\Omega_{\rm m}H_0^2} r(z_{\rm CMB}) , \\
\ell_{\rm a} &\equiv& \pi r(z_{\rm CMB})/r_{\rm s}(z_{\rm CMB}) ,
\end{eqnarray}
where $\Omega_{\rm m}$ and $H_0$ are the matter density parameter and
the Hubble parameter at present time, $r(z_{\rm CMB})$ is the comoving
distance from observer to the redshift of decoupling $z_{\rm CMB}$, and
$r_{\rm s}(z_{\rm CMB})$ is the sound horizon at the decoupling. We
calculate the redshift of decoupling via the fitting formula given in
\cite{Hu:1996}.

We estimate the 4$\times$4 covariance matrix for $R,~\ell_a,~\Omega_{\rm
b}h^2~{\rm and}~n_{\rm s}$; $\Omega_{\rm b}$ is the baryon density
parameter, $h$ is the normalized Hubble parameter, and $n_s$ is the
spectral index for the primordial power spectrum, from mock COrE data,
which is composed of the temperature, polarization and CMB lensing
potential angular power spectra.  We adopt the specification of the CMB
experiment following the COrE white paper\cite{COrE}, and we select the mean values of
WMAP7+BAO+$H_0$ analysis for $\Lambda$CDM model \cite{Komatsu:2010fb} as a fiducial
model of mock data.  In Table~\ref{tb:shift_core}, we show the mean
values and their r.m.s (Top) and the normalized covariance matrix for
$R,~\ell_a,~\Omega_{\rm b}h^2~{\rm and}~n_{\rm s}$ (Bottom), which are
estimated from our mock data for COrE, respectively.

Compared to the results of the Planck satellite given in
\cite{Mukherjee:2008}, we find that the correlation between $\Omega_{\rm
b}h^2$ and $\ell_a$ is reduced.  This is not only because COrE data
provide significantly tighter constraints, but attributable to our setup
of analysis. The damping of temperature fluctuations on small-scale
depends on $\Omega_{\rm b}h^2$, but the effects on this scale degenerate
with other cosmological parameters such as the primordial helium mass
fraction $Y_p$. In our analysis, we do not treat $Y_p$ as a free
parameter, and determine the value of $Y_p$ from baryon density
parameter $\Omega_{\rm b}h^2$ because $Y_p$ can be predicted from Big
Bang Nucleosynthesis as a function of baryon and radiation
densities\cite{Hamann:2008}. Therefore, the high-quality data on small
scales by COrE have potential to put tighter constraint on $\Omega_{\rm
b}h^2$.

\begin{table}[t]
\begin{center}

{The mean value and their r.m.s variance}\\
\vspace{1mm}
\begin{tabular}{ccc}
\hline \hline
\makebox[4cm]{Parameter} & 
\makebox[4cm]{Mean value} & 
\makebox[4cm]{r.m.s variance} \\
\hline
 $R$ & 
1.734  &  0.001617 \\
$\ell_a$ & 
302.1  &  0.02274  \\
$\Omega_{\rm b}h^2$ & 
0.02250   &  0.00005095 \\
$n_{\rm s}$ &
0.9615  &  0.001898 \\
\hline \hline
\end{tabular}

\vspace{5mm}
{The normalized covariance matrix}\\
\vspace{1mm}
\begin{tabular}{ccccc}
\hline \hline
 & 
\makebox[3cm]{$R$} & 
\makebox[3cm]{$\ell_a$} & 
\makebox[3cm]{$\Omega_{\rm b}h^2$} & 
\makebox[3cm]{$n_{\rm s}$} \\
\hline
 $R$ & 
 1.000000    &    0.418460   &    -0.347250   &    -0.542950 \\
$\ell_a$ & 
 0.418460    &    1.000000   &    -0.025810   &    -0.308290 \\
$\Omega_{\rm b}h^2$ & 
-0.347250    &   -0.025810   &     1.000000   &    -0.196580 \\
$n_{\rm s}$ &
-0.542950    &   -0.308290   &    -0.196580   &     1.000000 \\
\hline \hline
\end{tabular}

\end{center}
\caption{The mean values and rms variances for $R,~\ell_a,~\Omega_{\rm
b}h^2,~n_{\rm s}$ (Top) and their covariance matrix (Bottom), which are
estimated from the mock COrE data.  } \label{tb:shift_core}
\end{table}

\subsection{BAO}

Baryon acoustic oscillation (BAO) can be used as a geometrical measure
of the cosmic expansion, which can constrain the nature of dark energy.
Its characteristic scale is set by the sound horizon, which is the
distance traveled by the acoustic waves in baryon-photon plasma by the
time of recombination. Such characteristic scale can be measured
oriented along the line-of-sight and in its transverse direction, which
can probe the Hubble parameter and the angular diameter distance,
respectively. Thus we make use of the following quantities as a
geometrical measure:
\begin{equation}
x_1 = \frac{d_A (z)}{r_s},
\qquad
x_2 = H(z) r_s
\end{equation}
where $d_A (z)$ is the comoving angular diameter distance and $r_s$ is
the sound horizon at the baryon-drag epoch.  $H(z)$ is the Hubble
expansion rate at a redshift $z$.  To obtain an expected constraint in
future BAO observations, we assume predicted fractional uncertainties in
the above quantities $x_1$ and $x_2$ for bigBOSS \cite{Schlegel:2011zz}.

\subsection{Type-Ia Supernovae}
\label{sec:SNe}

The observables from a type-Ia supernovae survey are apparent magnitudes
$m_i$, and provide the measurements of luminosity distances $d_L(z_i)$
through the distance modulus as
\begin{equation}
\mu(z_i) \equiv m_i - M = 5\log_{10} d_{L}(z_i) + 25 , 
\end{equation}
where $M$ represents the absolute magnitude. 

In the estimation using Bayesian analysis, we can use the likelihood
with several simplifying assumptions for each model given as
\begin{equation}
        \ln {\cal L}_{(\rm SNe)} \simeq \frac{1}{2} \chi^2 _{(\rm SNe)}({\bm \mu_{i,{\rm mod}}})
        = \frac{1}{2} \sum_{i} \frac{ [ \mu_{{\rm fid},i} - \mu(z_i) ]^2}{\sigma^2_{i}} ,
\label{eq:chisq_SNe}
\end{equation}
where we assume some redshift bins and the index of summation $i$ runs
through all redshift bins, and $\mu_{{\rm fid},i}$ represents the
distance modulus of a fiducial model at $i$-th redshift bin and
$\sigma_i$ is its uncertainty.

We assume a Stage IV survey as described in the Dark Energy Task Force
report\cite{DETF} 
and the
statistical random error of observed apparent magnitude is described by
$\sigma_{\rm D}=0.1$. If we assume only the statistical random error and
there is no correlation of errors between different samples, the
uncertainty of the distance modulus at $i$-th redshift bin $\sigma_{i}$
can be written as
\begin{equation}
        \sigma_i = \sigma_{\rm D}/\sqrt{N_{{\rm bin}, i}}~, 
\end{equation}
where $N_{{\rm bin}, i}$ is the number of samples at $i$-th redshift
bin, and its concrete value is also given in \cite{DETF}.

\subsection{Constraints with background quantities}
\label{sec:const_b}

Now we investigate constraints only from background quantities, and
study the effects of a prior which forbids the phantom equation of
state, $w_X(a)<-1$, with the method described above.

\begin{figure}[t]
  \begin{center}
    \resizebox{170mm}{!}{ 
\includegraphics{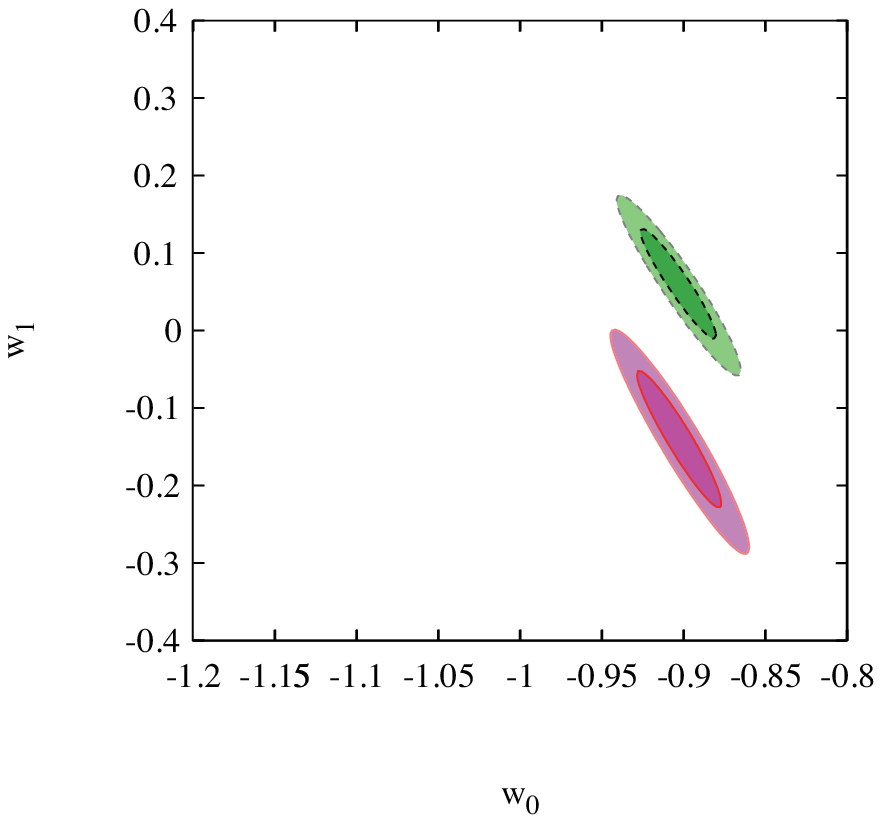}
  \hspace{-0.5cm}  
\includegraphics{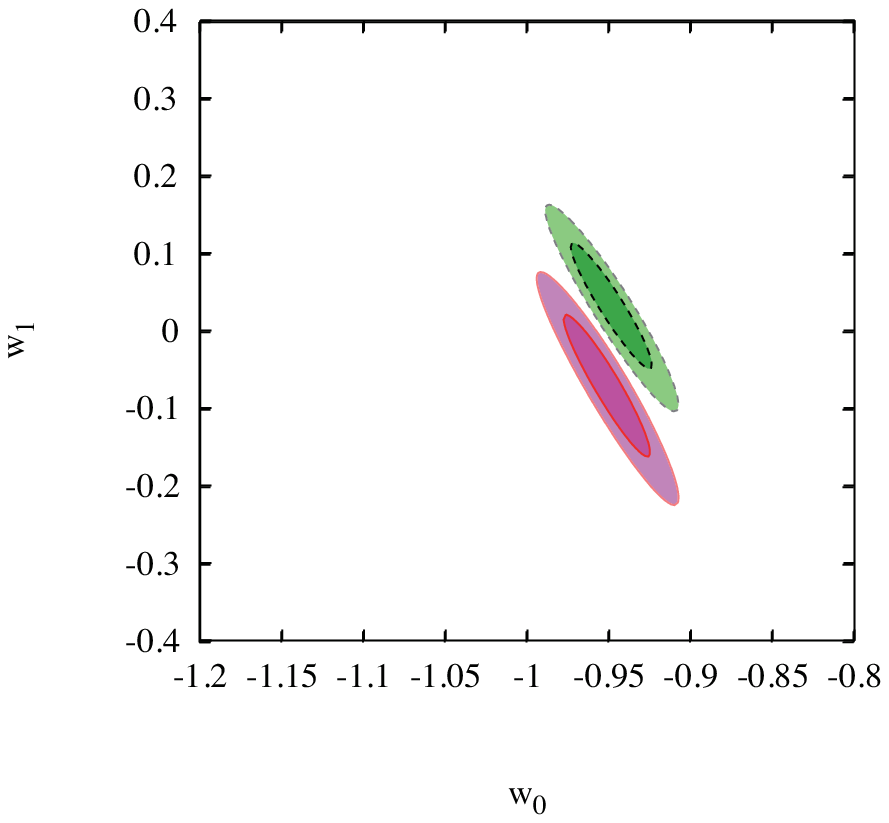}
}
  \end{center}
  \vspace{-0.5cm}
  \caption{Expected constraints for Parameterization I. The fiducial
  models are assumed to have $V(\phi) = (1/2) m^2 \phi^2$ (red/shaded)
  with model parameters being chosen such that the present equations of
  state become $w_X(a_0)=-0.9$ (left) and $-0.95$ (right), and $V(\phi)
  = \beta M_{\rm pl}^4 \left( \phi/ M_{\rm pl}\right)^\alpha$
  (green/shaded) with the present equations of state being
  $w_X(a_0)=-0.9$ (left) and $-0.95$ (right), which correspond to
  $\alpha=-1/3$ and $-1/6$, respectively.}
  \label{fig:bg_cpl}
\end{figure}

\begin{figure}[t]
  \begin{center}
    \resizebox{170mm}{!}{ 
\includegraphics{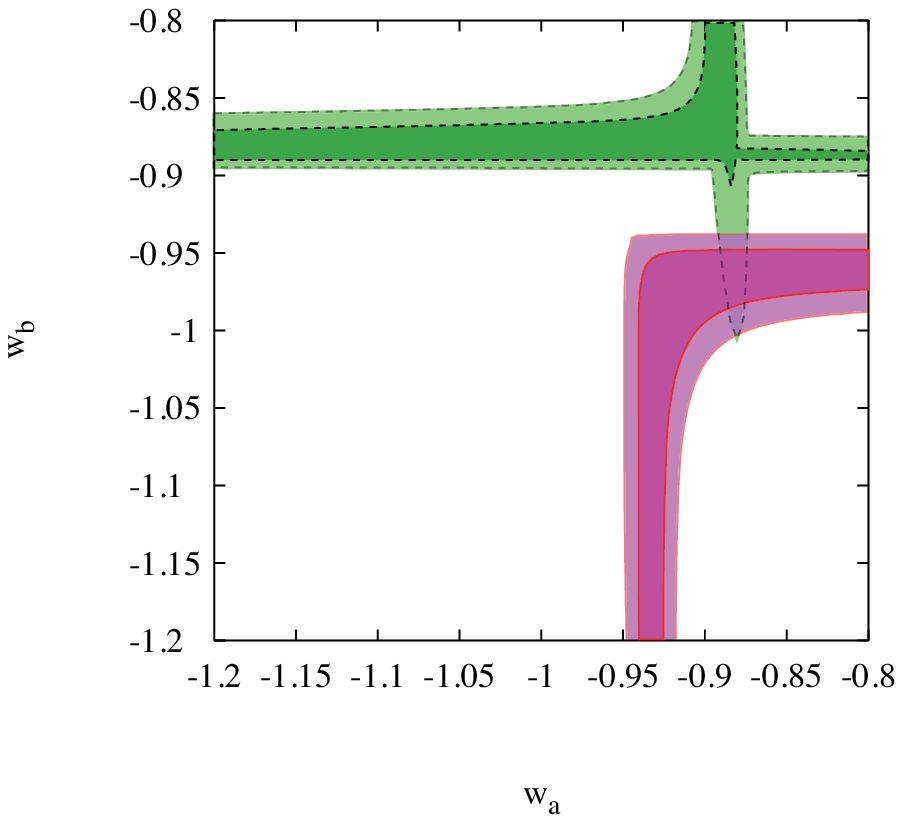}
  \hspace{-0.5cm}  
\includegraphics{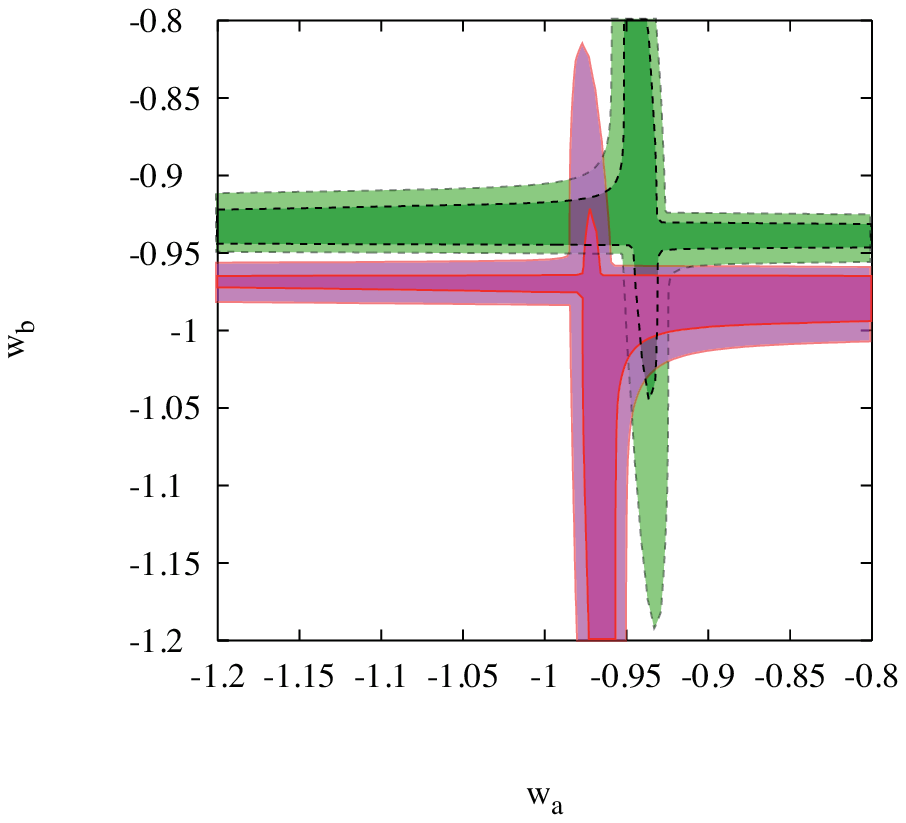}
}
  \end{center}
  \vspace{-0.5cm}
  \caption{Expected constraints for Parameterization II. The fiducial
  models are assumed to have $V(\phi) = (1/2) m^2 \phi^2$ (red/shaded) with
  model parameters being chosen such that the present equations of state
  become $w_X(a_0)=-0.9$ (left) and $-0.95$ (right), and $V(\phi) =
  \beta M_{\rm pl}^4 \left( \phi/ M_{\rm pl}\right)^\alpha$
  (green/shaded) with the present equations of state being $w_X(a_0)=-0.9$
  (left) and $-0.95$ (right), which correspond to $\alpha=-1/3$ and
  $-1/6$, respectively.
}
\label{fig:bg_hm}
\end{figure}

In Figures~\ref{fig:bg_cpl} and \ref{fig:bg_hm}, we show the constraints
for Parameterization I and Parameterization II, respectively, using the
thawing models of the potential $V(\phi)=(1/2)m^2\phi^2$ (red/shaded)
and the freezing model (green/shaded) as fiducial models.
 
We adopt the model parameters for the fiducial models such that the
present equations of state become $w_X(a_0)=-0.90$ and
$w_X(a_0)=-0.95$, and these values are based on the solutions of the
tracker model with $\alpha=-1/3$ and $\alpha=-1/6$, respectively. The
prior on the dark energy equation of state is not assumed in the figures. We
find that the thawing and the freezing models can be marginally
differentiated at $1\sigma$ CL for both Parameterizations as long as the
present equation of state $w_X \simeq -0.90$, which is less constrained in
comparison to the analysis done in the main text, including the
information of the fluctuations.

\begin{figure}[t]
  \begin{center}
    \resizebox{170mm}{!}{
\includegraphics{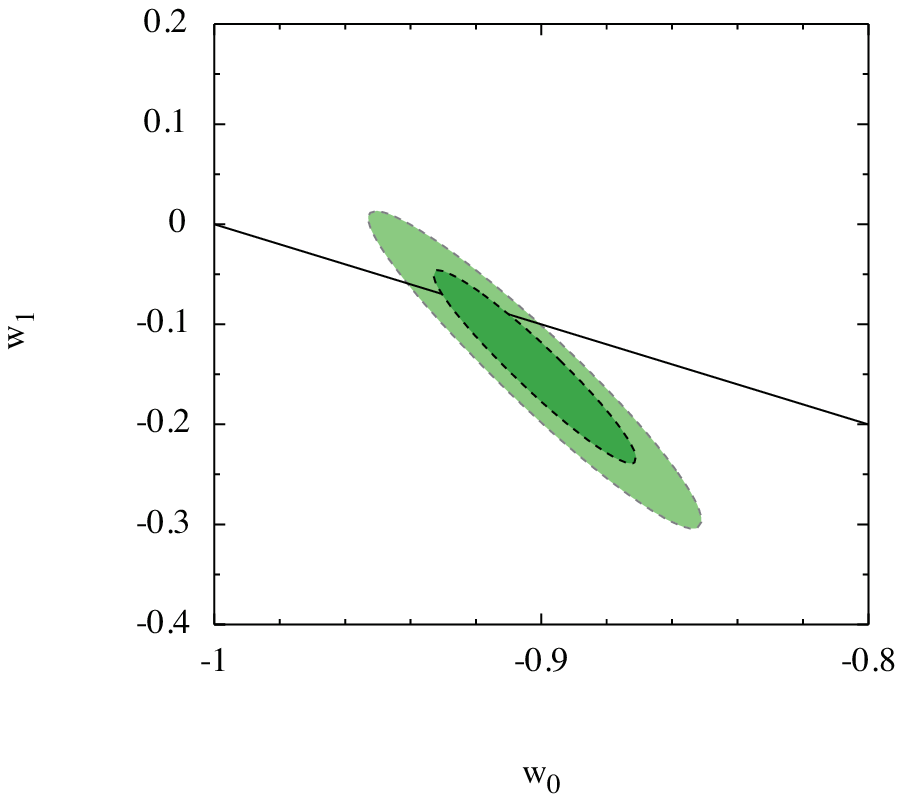}
\hspace{-1.5cm}   
\includegraphics{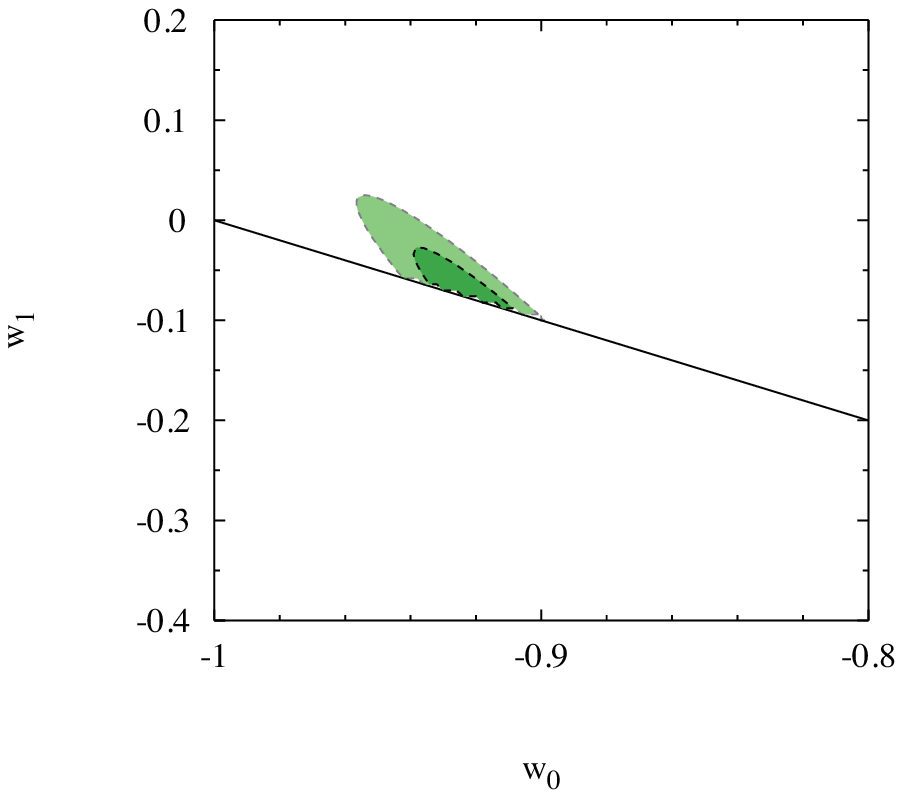}
} 
\end{center} 
\vspace{-0.5cm} 
  \caption{Expected constrains for Parameterization I. The fiducial
  model is assumed to have $V(\phi) = (1/2)m^2\phi^2$ with model
  parameters being chosen such that the present equation of state
  becomes $w_X(a_0)=-0.9$. The left panel shows the result with the
  equation of state being allowed to enter the phantom region,
  $w_X(a)<-1$, and the right one shows the result with the phantom
  equation of state, $w_X(a)<-1$, being prohibited.}
  \label{fig:cpl_phi2}
\end{figure}

Finally, we show the constraints from CMB (COrE), SNe (JDEM) and BAO
(bigBOSS) with Parameterization I to see the effects of a prior which
forbids the equation of state from crossing $w_X=-1$ in
Figure~\ref{fig:cpl_phi2}.  The effects of a prior do not seem sensitive
to the result on these parameter spaces so much, but it shifts the
center value of constraint slightly.

\clearpage



\begin{thebibliography}{100}


\bibitem{quint}

Y.~{Fujii}.
\newblock {\em \prd} {\bf 26}, 2580--2588, (1982);
%
R.~D. {Peccei}, J.~{Sol{\`a}}, and C.~{Wetterich}.
\newblock {\em Physics Letters B}, {\bf 195}, 183--190, (1987);
%
L.~H. {Ford}.
\newblock {\em \prd} {\bf 35}, 2339--2344, (1987);
%
C.~{Wetterich}.
\newblock {\em Nuclear Physics B}, {\bf 302}, 668--696, (1988).

\bibitem{quint2}

B.~{Ratra} and P.~J.~E. {Peebles}.
\newblock {\em \prd} {\bf 37}, 3406--3427, (1988);
  P.~J.~E.~Peebles and B.~Ratra,
  Astrophys.\ J.\  {\bf 325}, L17 (1988).

\bibitem{quint3}

Y.~{Fujii} and T.~{Nishioka}.
\newblock {\em \prd} {\bf 42}, 361--370, (1990);
%
T.~{Chiba}, N.~{Sugiyama}, and T.~{Nakamura}.
\newblock {\em \mnras}, {\bf 289}, L5--L9, (1997)
\newblock [astro-ph/9704199];
%
P.~G. {Ferreira} and M.~{Joyce}.
\newblock {\em Physical Review Letters}, {\bf 79}, 4740--4743, (1997)
\newblock [astro-ph/9707286];
%
E.~J. {Copeland}, A.~R. {Liddle}, and D.~{Wands}.
\newblock {\em \prd} {\bf 57}, 4686--4690, (1998)
\newblock [gr-qc/9711068];
%
R.~R. {Caldwell}, R.~{Dave}, and P.~J. {Steinhardt}.
\newblock {\em Physical Review Letters}, {\bf 80}, 1582--1585, (1998)
\newblock [astro-ph/9708069].

\bibitem{Chiba:2000}
T.~{Chiba}, T.~{Okabe}, and M.~{Yamaguchi}.
\newblock {\em \prd} {\bf 62}(2), 023511, (2000)
\newblock [astro-ph/9912463];
%
C.~{Armendariz-Picon}, V.~{Mukhanov}, and P.~J. {Steinhardt}.
\newblock {\em Physical Review Letters}, {\bf 85}, 4438--4441, (2000)
\newblock [astro-ph/0004134];

\bibitem{Kamenshchik:2001}
A.~{Kamenshchik}, U.~{Moschella}, and V.~{Pasquier}.
\newblock {\em Physics Letters B}, {\bf 511}, 265--268, (2001).
\newblock [gr-qc/0103004].
%
M.~C. {Bento}, O.~{Bertolami}, and A.~A. {Sen}.
\newblock {\em \prd} {\bf 66}(4), 043507, (2002).
\newblock [gr-qc/0202064].

\bibitem{Caldwell:2005tm} 
  R.~R.~Caldwell and E.~V.~Linder,
  Phys.\ Rev.\ Lett.\  {\bf 95}, 141301 (2005)
  [astro-ph/0505494].

\bibitem{tracker_quin}

I.~{Zlatev}, L.~{Wang}, and P.~J. {Steinhardt}.
\newblock {\em Physical Review Letters}, {\bf 82}, 896--899, (1999)
\newblock [astro-ph/9807002];
%
P.~J. {Steinhardt}, L.~{Wang}, and I.~{Zlatev}.
\newblock {\em \prd} {\bf 59}(12), 123504, (1999)
\newblock [astro-ph/9812313].

\bibitem{tracker_quin_model}
P.~Binetruy, 
  Phys.\ Rev.\  D {\bf 60}, 063502 (1999)
  [arXiv:hep-ph/9810553];
  P.~Brax and J.~Martin, 
  Phys.\ Lett.\  B {\bf 468}, 40 (1999)
  [arXiv:astro-ph/9905040];
  A.~Masiero , M.~Pietroni and F.~Rosati, 
  Phys.\ Rev.\  D {\bf 61}, 023504 (2000)
  [arXiv:hep-ph/9905346];
 E.~J.~Copeland, N.~J.~Nunes and F.~Rosati, 
  Phys.\ Rev.\  D {\bf 62}, 123503 (2000)
  [arXiv:hep-ph/0005222].

\bibitem{pNGB_quin}

J.~A. {Frieman}, C.~T. {Hill}, A.~{Stebbins}, and I.~{Waga}.
\newblock {\em Physical Review Letters}, {\bf 75}, 2077--2080, (1995).
\newblock [astro-ph/9505060];
%
K.~{Choi}.
\newblock {\em \prd} {\bf 62}(4), 043509, (2000).
\newblock [hep-ph/9902292].
%
J.~E.~Kim,
  JHEP {\bf 0006}, 016 (2000)
  [arXiv:hep-ph/9907528];
 Y.~Nomura, T.~Watari and T.~Yanagida, 
  Phys.\ Rev.\  D {\bf 61}, 105007 (2000)
  [arXiv:hep-ph/9911324];
 J.~E.~Kim and H.~P.~Nilles,
  Phys.\ Lett.\  B {\bf 553}, 1 (2003)
  [arXiv:hep-ph/0210402].

\bibitem{Ringeval:2010hf} 
  C.~Ringeval, T.~Suyama, T.~Takahashi, M.~Yamaguchi and S.~Yokoyama,
  Phys.\ Rev.\ Lett.\  {\bf 105}, 121301 (2010)
  [arXiv:1006.0368 [astro-ph.CO]].

\bibitem{Ma:1995}
C.-P. {Ma} and E.~{Bertschinger}.
\newblock {\em \apj}, {\bf 455}, 7, (1995).
\newblock [arXiv:astro-ph/9506072].

\bibitem{Chevallier:2000qy}
  M.~Chevallier and D.~Polarski,
  Int.\ J.\ Mod.\ Phys.\ D {\bf 10}, 213 (2001)
  [arXiv:gr-qc/0009008].

\bibitem{Linder:2002et}
  E.~V.~Linder,
  Phys.\ Rev.\ Lett.\  {\bf 90}, 091301 (2003)
  [arXiv:astro-ph/0208512].

\bibitem{Huterer:2000mj}
  D.~Huterer and M.~S.~Turner,
  Phys.\ Rev.\ D {\bf 64}, 123527 (2001)
  [arXiv:astro-ph/0012510].

\bibitem{Weller:2001gf}
  J.~Weller and A.~Albrecht,
  Phys.\ Rev.\ D {\bf 65}, 103512 (2002)
  [arXiv:astro-ph/0106079].

\bibitem{Frampton:2002vv}
  P.~H.~Frampton and T.~Takahashi,
  Phys.\ Lett.\ B {\bf 557}, 135 (2003)
  [arXiv:astro-ph/0211544].
  
\bibitem{Hannestad:2004cb}
  S.~Hannestad, E.~Mortsell,
  JCAP {\bf 0409}, 001 (2004).
  [astro-ph/0407259].

\bibitem{Ichiki:2007}
K.~{Ichiki} and T.~{Takahashi}.
\newblock {\em \prd} {\bf 75}(12), 123002, (2007).
\newblock [arXiv:astro-ph/0703549].

\bibitem{Hu:1998}
  W.~Hu,
  \newblock {\em Astrophys. J.},  {\bf 506}, 485 (1998)
  \newblock [arXiv:astro-ph/9801234].

\bibitem{Ilic:2010}
S.~{Ili{\'c}}, M.~{Kunz}, A.~R. {Liddle}, and J.~A. {Frieman}.
\newblock {\em \prd} {\bf 81}(10), 103502, (2010).
\newblock [arXiv:1002.4196].

\bibitem{Chiba:2009}
T.~{Chiba}.
\newblock {\em \prd} {\bf 79}(8), 083517, (2009).
\newblock [arXiv:0902.4037].

\bibitem{Chiba:2010}
T.~{Chiba}, M.~{Siino}, and M.~{Yamaguchi}.
\newblock {\em \prd} {\bf 81}(8), 083530, (2010).
\newblock [arXiv:1002.2986].


\bibitem{Crittenden:2007}
R.~{Crittenden}, E.~{Majerotto}, and F.~{Piazza}.
\newblock {\em Physical Review Letters}, {\bf 98}(25), 251301, (2007).
\newblock [arXiv:astro-ph/0702003].

\bibitem{Linder:2006}
E.~V. {Linder}.
\newblock {\em \prd} {\bf 73}(6), 063010, (2006).
\newblock [arXiv:astro-ph/0601052].

\bibitem{Sachs:1967}
R.~K. {Sachs} and A.~M. {Wolfe}.
\newblock {\em \apj}, {\bf 147}, 73, (1967).

\bibitem{Lewis:2006}
A.~{Lewis} and A.~{Challinor}.
\newblock {\em \physrep}, {\bf 429}, 1--65, (2006).
\newblock [arXiv:astro-ph/0601594].

\bibitem{Hu:2002}
W.~{Hu} and T.~{Okamoto}.
\newblock {\em \apj}, {\bf 574}, 566--574, (2002).
\newblock [arXiv:astro-ph/0111606].

\bibitem{Okamoto:2003}
T.~{Okamoto} and W.~{Hu}.
\newblock {\em \prd} {\bf 67}(8), 083002, (2003).
\newblock [arXiv:astro-ph/0301031].

\bibitem{Hirata:2003}
C.~M. {Hirata} and U.~{Seljak}.
\newblock {\em \prd} {\bf 67}(4), 043001, (2003).
\newblock [arXiv:astro-ph/0209489].

\bibitem{Lewis:2000}
A.~{Lewis}, A.~{Challinor}, and A.~{Lasenby}. 
  \newblock {\em \apj}, {\bf 538}, 473--476, (2000). 
  \newblock [arXiv:astro-ph/9911177]. 

\bibitem{Sheth:2001}
R.~K. {Sheth}, H.~J. {Mo}, and G.~{Tormen}.
\newblock {\em \mnras}, {\bf 323}, 1--12, (2001).
\newblock [arXiv:astro-ph/9907024].

\bibitem{Warren:2006}
M.~S. {Warren}, K.~{Abazajian}, D.~E. {Holz}, and L.~{Teodoro}.
\newblock {\em \apj}, {\bf 646}, 881--885, (2006).
\newblock [arXiv:astro-ph/0506395].

\bibitem{Bartelmann:2001}
M.~{Bartelmann} and P.~{Schneider}.
\newblock {\em \physrep}, {\bf 340}, 291--472, (2001).
\newblock [arXiv:astro-ph/9912508].

\bibitem{Ma:2006}
Z.~{Ma}, W.~{Hu}, and D.~{Huterer}.
\newblock {\em \apj}, {\bf 636}, 21--29, (2006).
\newblock [arXiv:astro-ph/0506614].


\bibitem{Smail:1994}
I.~{Smail}, R.~S. {Ellis}, and M.~J. {Fitchett}. 
  \newblock {\em \mnras} {\bf 270}, 245, (1994). 
  \newblock [arXiv:astro-ph/9402048].

\bibitem{Amara:2007}
A.~{Amara} and A.~{R{\'e}fr{\'e}gier}, 
  \newblock {\em \mnras} {\bf 381}, 1018--1026, (2007).
  \newblock [arXiv:astro-ph/0610127].


\bibitem{dePutter:2010}
R.~{de Putter}, D.~{Huterer}, and E.~V. {Linder}.
\newblock {\em \prd} {\bf 81}(10), 103513, (2010).
\newblock [arXiv:1002.1311].

\bibitem{COrE}
{The COrE Collaboration}. 
\newblock [arXiv:1102.2181].

\bibitem{LSST}
{LSST Science Collaborations}. 
\newblock [arXiv:0912.0201].

\bibitem{Lewis:2002}
A.~{Lewis} and S.~{Bridle}.
\newblock {\em \prd} {\bf 66}(10), 103511, (2002).
\newblock [arXiv:astro-ph/0205436].

\bibitem{Chiba:2013}
T.~{Chiba}, A.~{De Felice}, and S.~{Tsujikawa}.
\newblock {\em \prd} {\bf 87}(8), 083505, (2013).
\newblock [arXiv:1210.3859].

\bibitem{Dutta:2008}
S.~{Dutta} and R.~J. {Scherrer}.
\newblock {\em \prd} {\bf 78}(12), 123525, (2008).
\newblock [arXiv:0809.4441].

\bibitem{Euclid}
R.~{Laureijs}, J.~{Amiaux}, S.~{Arduini}, J.~. {Augu{\`e}res}, J.~{Brinchmann},
  R.~{Cole}, M.~{Cropper}, C.~{Dabin}, L.~{Duvet}, A.~{Ealet}, and et~al.
  \newblock [arXiv:1110.3193]. 

\bibitem{Huterer:06}
D.~{Huterer}, M.~{Takada}, G.~{Bernstein}, and B.~{Jain}.
\newblock {\em \mnras}, {\bf 366}, 101--114, (2006).
\newblock [arXiv:astro-ph/0506030].

\bibitem{Das:2011}
S.~{Das}, R.~{de Putter}, E.~V. {Linder}, and R.~{Nakajima}.
\newblock {\em ArXiv e-prints}, 1102.5090, (2011).
\newblock [arXiv:1102.5090].

\bibitem{DETF}
A.~{Albrecht}, G.~{Bernstein}, R.~{Cahn}, W.~L. {Freedman}, J.~{Hewitt},
  W.~{Hu}, J.~{Huth}, M.~{Kamionkowski}, E.~W. {Kolb}, L.~{Knox}, J.~C.
  {Mather}, S.~{Staggs}, and N.~B. {Suntzeff}.
\newblock {\em ArXiv Astrophysics e-prints}, arXiv:astro-ph/0609591, (2006).
\newblock [arXiv:astro-ph/0609591].

\bibitem{Schlegel:2011zz} 
  D.~Schlegel {\it et al.}  [BigBoss Experiment Collaboration],
  arXiv:1106.1706 [astro-ph.IM].

\bibitem{Mukherjee:2008}
P.~{Mukherjee}, M.~{Kunz}, D.~{Parkinson}, and Y.~{Wang}.
\newblock {\em \prd} {\bf 78}(8), 083529, (2008).
\newblock [arXiv:0803.1616].

\bibitem{Hu:1996}
W.~{Hu} and N.~{Sugiyama}.
\newblock {\em \apj}, {\bf 471}, 542, (1996).
\newblock [arXiv:astro-ph/9510117].

\bibitem{Komatsu:2010fb}
  E.~Komatsu {\it et al.},
  arXiv:1001.4538 [astro-ph.CO].

\bibitem{Hamann:2008}
J.~{Hamann}, J.~{Lesgourgues}, and G.~{Mangano}.
\newblock {\em \jcap}, {\bf 3}, 4, (2008).
\newblock [arXiv:0712.2826].

\bibitem{Mukherjee:08}
P.~{Mukherjee}, M.~{Kunz}, D.~{Parkinson}, and Y.~{Wang}.
\newblock {\em \prd} {\bf 78}(8), 083529, (2008).
\newblock [arXiv:0803.1616].

\end{thebibliography}
\end{document}